\RequirePackage{fix-cm}
\documentclass[smallextended]{svjour3}       
\smartqed  
\usepackage{graphicx}
\newcommand{\rem}[1]{}

\renewcommand\refname{}

\usepackage{natbib}
\setcitestyle{authoryear, open={(},close={)}}

\usepackage[utf8]{inputenc}
\usepackage[T1]{fontenc}

 \usepackage{color}

\newcommand{\rev}[1]{\textcolor{black}{#1}}

\begin{document}

\title{An Argument Communication Model of Polarization and Ideological Alignment\thanks{This project has received funding from the European Union’s Horizon 2020 research and innovation programme under grant agreement No 732942 (Opinion Dynamics and Cultural Conflict in European Spaces -- www.\textsc{Odycceus}.eu).}
}


\author{Sven Banisch         \and
        Eckehard Olbrich 
}


\institute{S. Banisch \& E. Olbrich\at
              Max Planck Institute for Mathematics in the Sciences\\
              Inselstraße 22\\
              D-04103 Leipzig\\
              Germany
               \\
              Tel.: +49-341-9959536\\
              \email{sven.banisch@UniVerseCity.de}           
}

\date{ }

\maketitle

\begin{abstract}
A multi-level model of opinion formation is presented which takes into account that attitudes on different issues are usually not independent.
In the model, agents exchange beliefs regarding a series of facts.
A cognitive structure of evaluative associations links different (partially overlapping) sets of facts to different political issues and determines an agents' attitudinal positions in a way borrowed from expectancy value theory.
If agents preferentially interact with other agents that hold similar attitudes on one or several issues, this leads to biased argument pools and increasing polarization in the sense that groups of agents selectively believe in distinct subsets of facts.
Besides the emergence of a bi-modal distribution of opinions on single issues that most previous opinion polarization models address, our model also accounts for the alignment of attitudes across several issues along ideological dimensions.
\keywords{Opinion Dynamics \and Group Polarization \and Ideological Consistency \and Belief Systems \and Cognitive-Evaluative Maps \and Argument Exchange}
\end{abstract}

\section{Introduction}
\label{sec:Einleitung}

Modeling the dynamics of political opinions is a complicated issue.
It is the interplay of a manifold of psychological and social processes that leads to the formation of complex landscapes of political preferences.
Many different theoretical pieces must be brought together for an encompassing understanding of political opinion dynamics and a single article like this can only address a few of them.

A particular focus of this article is on the multilevel structure of political opinion spaces.
On the one hand side, we distinguish arguments and beliefs about facts from a political issue on which an attitude is formed by considering that a whole series of facts speaking in favor or in disfavor of a certain position is usually involved in an attitudinal judgement.
In line with expectency-value theories of attitude research and measurement \citep{Fishbein1962ab,Fishbein1963investigation,Ajzen2001nature}, we assume that an attitude regarding a political issue is formed through a structure of evaluations regarding the different aspects that are relevant for the topic.
Also communication about those topics makes reference to these underlying argumentative dimensions.
On the other hand, we present a model in which different political issues may be discussed at the same time.
These topics are not independent but related through the underlying cognitive-evaluative meaning structures.
\rev{
With that, our model formalizes inter--attitudinal structures on the basis of the underlying thematic consistency or inconsistency which is closely related to the classical meaning of >>ideology<< (e.g. \citet[281]{Eagly1998attitude} or \citet[8]{Converse1964nature}).
}
Depending on the nature of the relations in these evaluative schemata, different patterns of political preferences evolve that lead to an organization along politico-ideological dimensions such as left versus right or liberal versus conservative.

The model proposed in this paper is an extension of the >>argument communication theory of bi-polarization (ACTB)<< that has been proposed in \cite{Maes2013differentiation}.
The ACTB attempts to explain the emergence of a bi-modal opinion distribution in the sense that small initial opinion differences are amplified in processes of social influence so that two antagonistic groups covering the extremes of an opinion spectrum emerge.
The argument-based approach by \cite{Maes2013differentiation} combines ideas from persuasive argument theory of group polarization \citep{Burnstein1975person,Burnstein1977persuasive,Isenberg1986group} with the assumption that homophily with respect to the opinions \citep{Byrne1961interpersonal,Huston1978interpersonal,McPherson2001birds,Wimmer2010beyond,Bakshy2015exposure} guides interaction and communication behavior.
Opinions are based on a series of pro- and con-arguments that are exchanged in an interaction process.
\rev{
While the argument exchange process affects believes at the lower level of facts, attitude or value homophily \citep{Lazarsfeld1954friendship,Byrne1961interpersonal} acts at the aggregate level of attitudes.
}
As a consequence, homophily creates a tendency that actors interact with like-minded others and therefore are likely to be exposed to arguments that further support their current attitudinal inclination.
Analogous to the >>law of group polarization<< described in \cite{Sunstein2002law}, this leads to biased argument pools that may reinforce group opinions in the direction of an initial inclination.
Apart from the social bias of homophily no further psychological biases are needed to explain the emergence of a bi-polar opinion distribution.
\rev{
Corresponding to the experimental finding that deliberation has a larger effect for less salient topics \citep{Farrar2010disaggregating}, one could argue that models implementing ACTB are describing a situation at the beginning of a discourse, because at later stages all arguments might be already known to the participants.
}

ACTB as well as quite a series of other recent polarization models focus on the formation of a bi-modal distribution of opinions regarding a single issue.
However, such strong patterns of opinion divergence with respect to single issues cover only one aspect of polarization \citep{DiMaggio1996have,Bramson2016disambiguation}.
Moreover, it has been empirically found only for certain morally charged >>polarizing<< topics such as abortion in US public opinion \citep{DiMaggio1996have}.
Besides different forms of social polarization such as, for instance, increasingly antagonistic references between two groups of different identity \citep{Uitermark2016dissecting,Mason2018uncivil}, also more specific notions of issue polarization have been put forth \citep{DiMaggio1996have,Bramson2016disambiguation}.
The sorting of opinions regarding diverse sets of political goals along ideological dimensions as recently identified in opinion data on American public opinion \citep{Dimock2014political} is of particular relevance in our context.
In fact, an organization of the complex landscapes of political preferences along a few axes such as left versus right, liberal versus conservative or technocratic versus ecological \citep{Leuthold2007making} can only acquire meaning if political preferences regarding a multitude of political goals are correlated over a population and constrained in form of a belief system or ideology \citep{Converse1964nature}.

The main aim of the paper is to show that ACTB is easily extended to account for this kind of opinion sorting or ideological alignment as well.
\rev{
For this purpose, we extend the model to multiple issues which are cognitively related because some arguments are relevant for more than one issue.
While a series of multi-dimensional models of opinion formation that explain persistent opinion plurality is available in the literature \citep{Axelrod1997dissemination,Macy2003polarization,Urbig2005dynamics,Fortunato2005vector,Baldassarri2007dynamics,Lorenz2008fostering,Huet2008rejection,Flache2011small}, the interrelatedness of opinions on various issues has not been addressed within these models.
}
\rev{
We think that the incorporation of issue interdependence on the basis of underlying arguments can contribute} to a better understanding of the formation of complex but at the same time specifically organized opinion landscapes.
For this purpose we assume the existence of cognitive-evaluative maps \citep[cf. e.g.][214]{Rosa2016Resonanz} which encode the evaluative meaning of different beliefs with respect to the different issues of discussion.
To simplify the interpretation of the dynamical behavior of the model, we consider that these evaluation structures have been acquired in the same socio-cultural context and are collectively shared by all agents.
\rev{At the same time,} this can be seen as a way to incorporate some notion of cultural meaning \citep{Berger1970gesellschaftliche,Schuetz2017strukturen} into models of opinion dynamics, and we will discuss the possibility that different >>cultures<< with their specific evaluative schemata engage in an argument exchange process and relate this to recent approaches to infer shared belief systems from attitudinal data \citep{Baldassarri2014ideologies, Daenekindt2017how}.
Our structure is inspired by structural theories of attitudes \citep{Fishbein1962ab,Fishbein1963investigation,Ajzen2001nature} in which a systematic distinction between beliefs and evaluations is made.
In our model we consider the evaluations as externally given and model the exchange of arguments about a set of beliefs that are relevant in the thematic complex at question.
As the same beliefs may contribute positively or negatively to different issues, the evaluative structure already imposes constraints on the admissible positions in the opinion space.
Only certain combinations of opinions are possible and the argument exchange process of the ACTB can inform us about the likeliness of specific configurations especially in conditions of polarization.

\rev{
In this paper, we concentrate on the capacity of the model to account for the emergence of coherent bundles of opinions.
This is motivated by the fact that -- implicitly or explicitly -- spatial theories of political opinion and voting \citep{Downs1957economic,Leuthold2007making,Laver2014measuring} rely on such an ideological ordering of political preferences, but the dynamical processes which may lead to such issue alignments have not yet been addressed in the opinion dynamics literature.
We propose a model that explains opinion alignment on multiple issues on the basis of the argumentative interrelatedness of different issues and the cognitive constraints this imposes.
To analyze the principal effects that structured attitudes can bring about, we perform a series of simulation experiments for different prototypic settings in the case of two issues (Sec. \ref{sec:sim2issues}).
As attitudes in two-issue models are distributed in a two-dimensional opinion space (two judgements on a seven-point scale ranging from -3 (extreme disfavor) to +3 (extreme favor) in our cases) there are different possibilities to define >>opinion distance<< and consequently homophily.
We explore four different modes including the case where different opinions with respect to a single >>ideologically loaded<< issue determines if arguments are adopted from an interaction partner or not.
Also the case that certain beliefs operate as >>identity signals<< \citep{Bacharach2001trustsigns} indicating to which ideological subgroup one belongs is considered.
In this way, we obtain a systematic picture of the kind of opinion correlations that may be induced between two issues that are compatible or incompatible in the light of a series of facts.  
}

\rev{
At the same time, however, the model is part of a larger research agenda that aims to address real debates around specific topics using opinion dynamics approaches.
We provide an example application related to climate change and electricity production in Section \ref{sec:threeissueexample} to illustrate this potential of linking opinion dynamics more closely to empirical data on political opinion.
This shows that ACTB is a very useful starting point for developing models that go beyond the reproduction of stylized facts for it allows to represent the content dimensions and arguments of real debates.
Addressing the question of issue alignment with reference to the underlying argumentative dimensions, allows to relate coherent patterns of opinions to the ways different actors talk about the issues.
Recent advances in precision language processing \citep{Steels2017basics,VanEecke2018exploring} and argument mining \citep{Lippi2016argumentation} will afford new opportunities to develop empirically-grounded scenarios in which model assumptions can be tested.
Moreover, the structural and multilevel conception of opinions is generally compatible with common social science survey methods and closely resembles recent experimental designs to assess persuasiveness and effects of arguments \citep{Kobayashi2016relational,Shamon2019arguments}.
The proposed framework, therefore, paves the way for completely new ways of model validation and micro foundation which have been repeatedly identified as the most important frontiers in model-based research on opinion dynamics \citep{Sobkowicz2009modelling,Flache2017models}.
}

\rem{
and allow an elaboration of specific settings that can be related to real debates.
It is based on a structural conception of opinion that closely resembles attitudes 
First, as our example shows, 
rich enough to represent arguments used in real debates while being 
(i.) with its close relation to structural theories of attitudes, it is well-atuned to 
}


The paper is organized as follows.
We first introduce the model in Section \ref{sec:model}.
In Section \ref{sec:sim2issues} we analyze its basic behavior by looking at 12 cases differing in the type of evaluative structure and homophily mechanism for two issues.
Two cases are explored with some more detail in this section.
In Section \ref{sec:threeissueexample} we look at an example with three issues that illustrates how meaningful arguments can in principle be integrated into the structure.
We discuss the model from a broader perspective in Section \ref{sec:discussion} and draw a conclusion on this paper in Section \ref{sec:conclusion}.
\rev{
Notice that an online implementation of the model accompanies the paper and is briefly described in the end of Section \ref{sec:settings}.
}

\section{The Model} 
\label{sec:model}

We model a population of agents that exchange arguments about different political issues.
Our model is based on three different ingredients: (1) agents exchange their beliefs regarding a series of facts; (2) different (partially overlapping) sets of facts are associated with different political issues and an agent's attitude towards these issues is a function of the evaluative relevance of the facts for the different issues; and (3) agents preferentially interact with other agents that hold similar attitudes on the issues.
These combined processes give rise to polarization and >>constraint[s] or functional interdependenc[ies]<< \citep[cf.][3]{Converse1964nature} in the configuration of attitudes towards multiple political issues.

\textbf{Argument Strings.}
Consider a population of $N$ agents that exchange arguments about different political issues.
There are $N_A$ argument dimensions related to facts which an agent (say $s$) either believes to be true $a_{sk} = 1$ or not $a_{sk} = 0$.
That is, each agent holds a binary string $\vec{a}_s \in \{0,1\}^{N_A}$ representing her current beliefs in a number of facts.
\rev{
Within our setting argument communication corresponds to the exchange of beliefs in facts, and we do not further distinguish between the two.
}
As a convention, we shall index the argument dimensions by $k$ ($1 \leq k \leq N_A$) and the agents by $s$ and $r$ for sender and receiver ($1 \leq s,r \leq N$).
Consequently, the argument strings of the entire population can be represented by an $N \times N_A$ matrix $A$ in which single rows represent the argument strings of the agents and the element $a_{sk}$ denotes the belief of agent $s$ with respect to the $k$th factual dimension.

\begin{figure}[ht]
 \centering
 \includegraphics[width=0.7\linewidth]{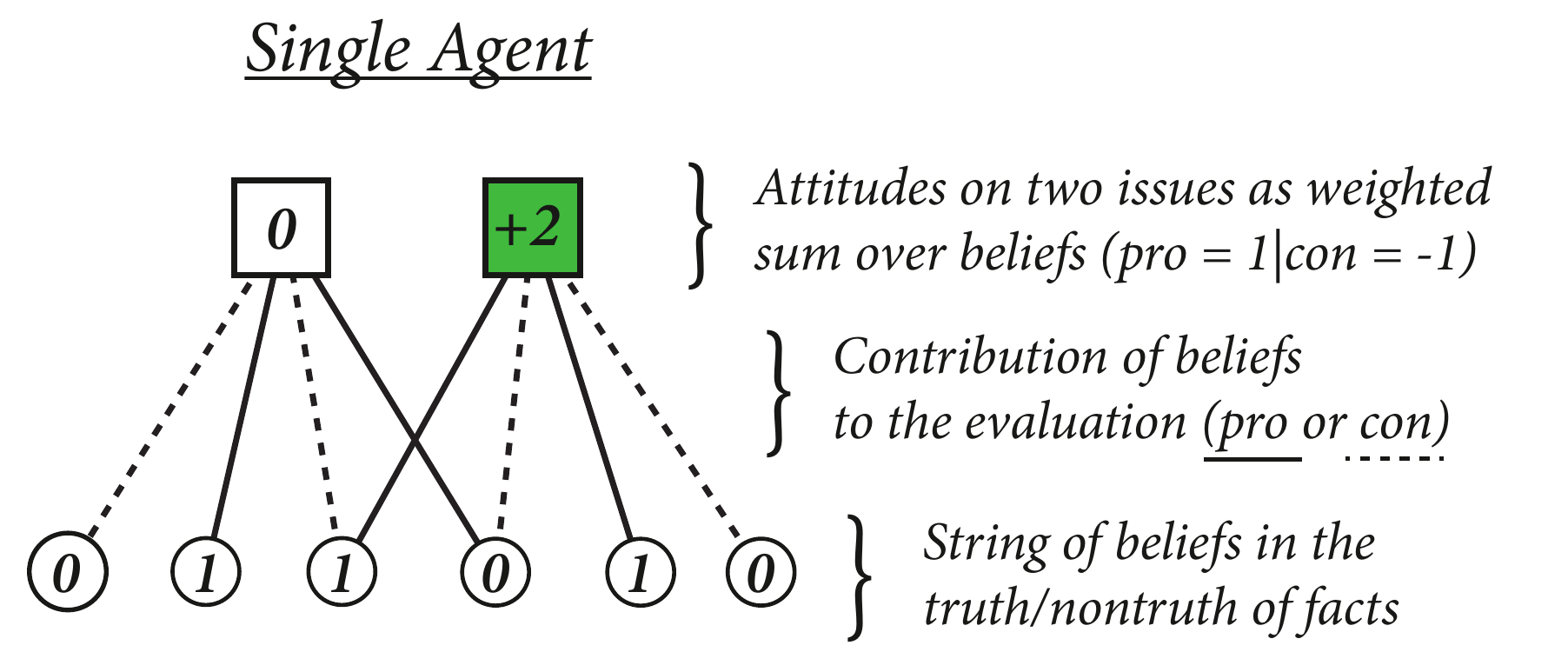}
 \caption{\rev{Cognitive architecture of the agents. Agents form attitudes on two different issues based on their beliefs in a number of facts which may contribute positively (pro, solid lines) or negatively (contra, dashed lines) to the attitudes.}}
 \label{fig:CognitiveArchitecture}
 \end{figure}

\textbf{Evaluative Structure.}
We consider that opinions are multi-level constructs.
Agents hold beliefs regarding a series of factual dimensions (encoded in $A$) and opinions on the set of issues are determined by specific configurations of beliefs.
One of the main assumptions of this work is that the link from beliefs to attitudes is realized by a cognitive-evaluative map that encodes how different factual dimensions contribute to an attitudinal judgement.
This map is modeled as a bi-partite graph $(\mathcal{I},\mathcal{A},\mathcal{C})$  which represents the relation between a set of issues $\mathcal{I}$ and the set of argumentative dimensions $\mathcal{A}$ related to facts.
We denote the cardinality of these two sets as $N_I$ and $N_A$ respectively.
The $N_A \times N_I$ evaluation matrix $C$ assigns values $c_{ki} \in [-1,1]$ to the set $\mathcal{C}$ of associations which represent (i.) whether an attitude object $i$ (a political issue in our case) is positively or negatively evaluated if a certain fact $k$ is believed to be true (sign$ (c_{ki})$), and (ii.) the extent to which that fact $k$ contributes to the evaluation of $i$ ($|c_{ki}|$).
This structure is inspired by the expectancy value model of attitudes following \cite{Fishbein1962ab,Fishbein1963investigation} and one may think of the set of factual beliefs as beliefs on the presence or absence of attributes.
\rev{
The resulting cognitive architecture is illustrated in Fig. \ref{fig:CognitiveArchitecture} for two issues (squares) and 6 facts (circles).
The evaluative structure contains four positive links (solid lines) and four negative links (dashed lines).
Notice that the third and forth factual dimensions are relevant to both issues and speak in favor of one but contra the other issue.
}

In the computation of the evaluative judgement -- that is, the attitude -- regarding the different issues, we follow the algebraic model of expectancy value theory in a straightforward manner.
Namely, the single argument-evaluation contributions are additively combined in the determination of the valence (degree of favor/disfavor) assigned the issue $i$:
\begin{equation}
o_s(i) = \sum\limits_{k=1}^{N_A} a_{sk} c_{ki}.
\label{eq:EvalIssue}
\end{equation}
Consequently, the attitudes on a whole set of $N_I$ issues for a single agent $s$ are given by the product $o_s = \vec{a}_s \cdot C$.
For the entire population we can then write in matrix form
\begin{equation}
O = A \cdot C.
\label{eq:EvalProjection}
\end{equation}
The element $o_{si} = o_s(i)$ in the resulting $N \times N_I$ matrix $O$ represents the attitude or opinion of agent $s$ towards issue $i$.
\rev{
Consider the agent shown in Fig. \ref{fig:CognitiveArchitecture} as an example.
It has a neutral attitude with respect to the first issue as one belief (2) supports a positive judgement, another belief (3) a negative one, and the third belief (5) is not relevant.
The attitude regarding the second issue is rather positive because two beliefs support a positive stance.
}

Generally, the evaluative structure $C$ may vary across individuals and every individual $s$ could be represented by its own structure $C_s$.
This would allow for inter-individual differences regarding the interpretation and relevance of facts.
Furthermore, while we model an exchange of arguments and assume that agents understand them equally, political discourse is actually often about the meaning of concepts involved in argumentative statements.
That is, attitude change may actually often come about by changes in the perception, evaluative meaning and relevance of arguments. 
We shall discuss both issues in Section \ref{sec:discussion}.
As explained in the Introduction, in most parts of this paper, we assume a shared and time-homogeneous evaluative structure $C_s = C$ for all agents which is motivated by the fact that individuals within the same socio-cultural context have internalized similar evaluative schemata.
We show that attitudes may polarize along ideologically coherent lines even if agents interpret facts in the same way because groups of agents selectively belief in distinct subsets of facts.

\begin{figure}[ht]
 \centering
 \includegraphics[width=0.99\linewidth]{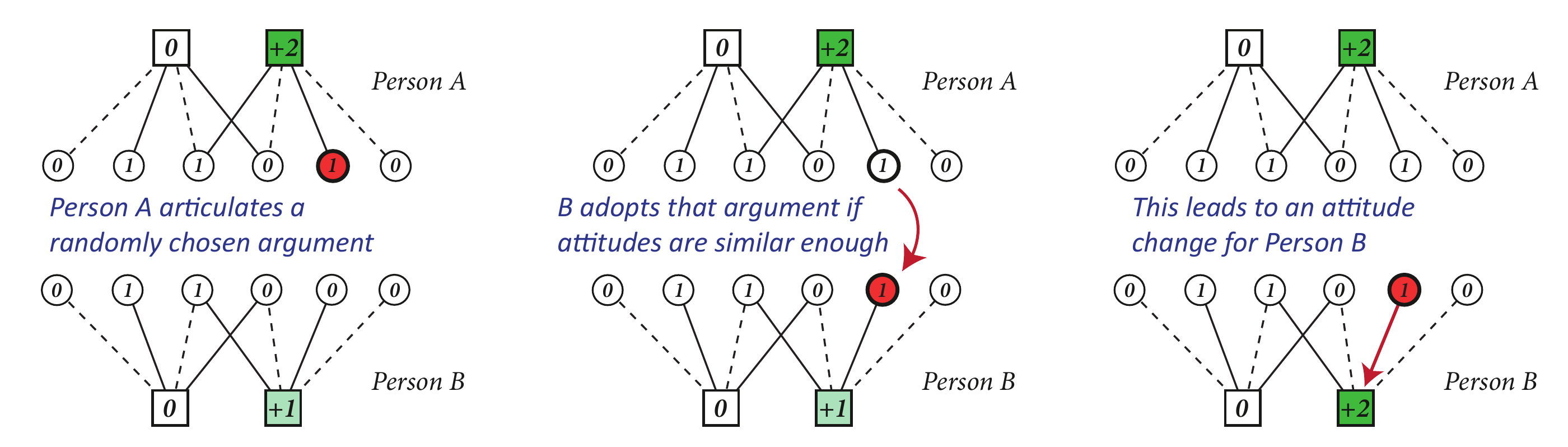}
 \caption{\rev{Summary of the argument exchange process performed during each step of the simulation. The sender (person A) chooses a random belief and presents it to its interlocutor. If the attitudes are similar enough (homophily, see below), the receiving agent (B) adopts that belief. As a result agent B has to recompute its attitudes (by (\ref{eq:EvalIssue})). In this example the agent received an argument relating positively to the second issue and a weakly positive opinion is further reinforced by this.}}
 \label{fig:ArgExchangeProcess}
 \end{figure}

\textbf{Argument Exchange.}
We assume a very simple mechanism for the argument exchange process \rev{and implement it as model of dyadic interaction.
When running the model an agent pair (sender $s$ and receiver $r$) with attitudes $o_s$ and $o_r$ is randomly selected at each time step.
They engage in argument exchange activity if their opinion vectors are similar enough as explained below.
The sender $s$, randomly choses an argumentative dimension $k$ and articulates her belief $a_k$ that fact $k$ is true or not.
}
The second agent $r$ receives that argument and adopts the respective belief, i.e. $a_{rk} = a_{sk}$.
With this simple procedure we follow previous approaches in opinion dynamics modeling with multidimensional opinions \citep{Axelrod1997dissemination,Banisch2010acs} and depart slightly from the more complex implementation in \cite{Maes2013differentiation} where the activation of one argument entails the deactivation of another.
The argument exchange mechanism is visually summarized in Fig. \ref{fig:ArgExchangeProcess}

\textbf{Bounded Confidence and Biased Argument Pools.}
Notice that in the argument exchange process implemented in the ACTB \citep{Maes2013differentiation} as well as in our implementation no forms of biased argument processing are integrated.
While the integration of such biases could be an interesting extension of the models, it is not needed to explain the emergence of bi-polar opinion distributions and the inter-attitudinal constraints that we address in this paper.
A >>homophily bias<< which creates a tendency that actors interact \rev{and exchange arguments} with like-minded others is sufficient. 

In our case, homophily is assumed to play out at the level of attitudes towards the $N_I$ issues and not on the level of the underlying arguments.
Besides ACTB \citep{Maes2013differentiation}, this idea has another important predecessor in the biological model of sympatric speciation developed by \cite{Kondrashov1986multilocus,Kondrashov1998origin} \rev{where the phenotypic expression of genes (and not the genes themselves) become functional in terms of natural selection.}
\rev{
In our context this means that attitudes (and not underlying cognitions) become functional regarding the selection of peers \citep[269]{Eagly1998attitude} such that agents with opposing attitudes do not engage in constructive interaction.
Just as phenotype assortativity leads to reproductively isolated gene pools and hence to the splitting of species \citep{Kondrashov1998origin}, homophily constraints at the attitude level lead to biased argument pools and attitude polarization in a sense closely related to \cite{Sunstein2002law}.
}


\rev{
There are different ways to integrate homophily assumptions into models of opinion dynamics.
While a threshold mechanism referred to as \emph{bounded confidence} is very common in continuous models of opinion dynamics \citep{Deffuant2000mixing,Hegselmann2002opinion,Fortunato2005vector,Urbig2005dynamics,Lorenz2007continuous2,Lorenz2008fostering}, a frequency-dependent interaction probability that takes into account the relative similarity of opinions with respect the entire popuation is used in \cite{Maes2013differentiation}.
In this work, we rely on the former and use the concept of bounded confidence as a simple mechanism for opinion homophily accounting for the well-established observation that people are more likely to interact if they hold similar views \citep{Byrne1961interpersonal,Huston1978interpersonal,McPherson2001birds}.
Others have motivated bounded confidence in terms of self--categorization assuming that people feel in a group with others who have similar opinions \citep{Lorenz2008fostering}.
Bounded confidence realizes this by a threshold $\beta$ on the opinion distance between two agents $s$ and $r$ such that no exchange takes place if $d(o_s, o_r) > \beta$ where $d(\ldotp,\ldotp)$ denotes some distance measure on the space of opinions.
}

\rev{
In our case, the position of agents in opinion space is determined by their belief strings $A$ and the attitude structure $C$ through (\ref{eq:EvalProjection}).
In a model aiming at describing the evolution of attitudes on various issues, several options for computing the distance become available.}
For instance, it might be that positions on one issue are much more salient with respect to the decision of whether or not to engage in communication with another agent.  
In this case, we take into account only the positions on this issue in the distance computation.
It might also be an asymmetric relation.
Furthermore, even the situation that a single argument signals an unacceptable stance of the interlocutor might be plausible in some cases. 
We will explore some of these options and provide an overview of their impact in the next section.

\rev{
Notice finally that the interaction behavior is only determined by opinion homophily, that is, by similarity in the attitude space \citep{Byrne1961interpersonal}.
We do not consider homophily related to status or socio--demographic characteristics such as age, gender or religion \citep{Lazarsfeld1954friendship} and the random pairing of agents is not mediated through a specific interaction network.
Demographic attributes and interaction structures have been integrated with ACTB in \cite{Maes2013short} to analyze the effect of demographic faultlines in group discussion processes.
In this paper, however, we focus on extending ACTB to multiple interrelated issues to understand how >>ideological faultlines<< may come about.
}


\section{Results: Two Issues}
\label{sec:sim2issues}

\subsection{Simulation Settings and Approach}
\label{sec:settings}

\begin{figure}[ht]
 \centering
 \includegraphics[width=0.69\linewidth]{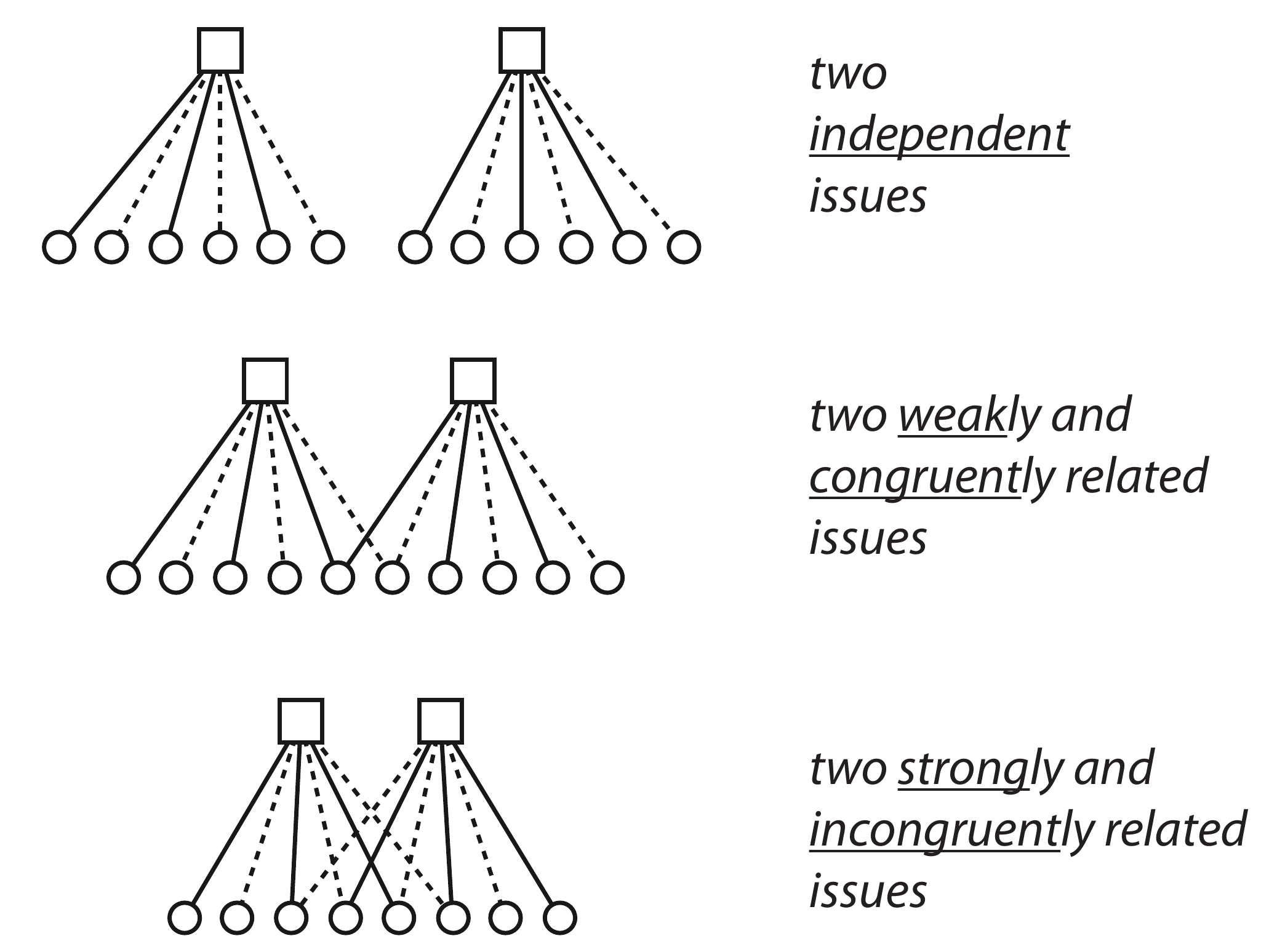}
 \caption{Three different evaluative maps are used that differ in the strength and congruency of the evaluative overlap. Solid lines indicate positive links and dashed line negative associations.}
 \label{fig:Settings}
 \end{figure}

\rev{
\textbf{Cognitive-Evaluative Maps.}
}
In this section we provide a general overview of the model behavior for the case of two issues.
All the results are based on simulations with $N=1000$ agents.
In all the cases we consider that 6 factual dimensions are relevant for each of the issues and that three of them contribute positively with $c_{ki} = 1$ and three negatively with $c_{ki} = -1$ to the respective attitude.
Following Eq. (\ref{eq:EvalProjection}), this means that the attitudinal judgements \rev{lie on a seven point scale} ranging from -3 (extreme disfavor) to +3 (extreme favor) and are neutral $o_i = 0$ if all facts are believed.
We look at three different conditions concerning the evaluative overlap with respect to the two issues:
\begin{enumerate}
\item 
the two issues are independent and there are 6 arguments relevant to the first and 6 other arguments relevant to the second issue,
\item
the two issues are weakly compatible in terms of the evaluative structure such that 2 factual dimensions contribute equally (one positively and one negatively) to the evaluation of the two issues,
\item
the two issues are strongly incompatible by assuming that 2 arguments contribute positively to the first and negatively to the second issue and another 2 arguments contribute negatively to the first and positively to the second. 
\end{enumerate}
The respective bi-partite graphs are shown in Fig. \ref{fig:Settings}.
Notice that there are 12 arguments in the first but only 8 in the last condition.
We have chosen this setup to make sure that the range of opinions is equal \rev{(seven point scale from -3 to +3)} in all the three conditions.

\rev{
\textbf{Modes of Homophily.}
}
Furthermore, we analyze the effect of four different homophily mechanisms by considering four different measures of distance upon which agents are assumed to judge whether or not they engage in effective communication with one another:
\begin{enumerate}
\item 
the Manhattan distance $d(o_s,o_r) = |o_s(1)-o_r(1)| + |o_s(2)-o_r(2)|$ taking into account the two issues,
\item
the Euclidean distance $d(o_s,o_r) = \sqrt{(o_s(1)-o_r(1))^2 + (o_s(2)-o_r(2))^2}$ that takes into account both issues as well,
\item
the distance (or opinion difference) with respect to one issue $i^*$, $d(o_s,o_r) = |o_s(i^*) - o_r(i^*)|$,
\item
and the difference with respect to a single belief $k^*$, $d(o_s,o_r) = |a_{sk^*} - a_{rk^*}|$.
\end{enumerate}
A summary of the 12 different combinations of homophily conditions and the three different evaluative structures is provided by Fig. \ref{fig:OverviewTable} (see Section \ref{sec:resultssummary}).

\rev{
\textbf{Initial Conditions.}
}
In all the simulations we consider throughout this paper, the population is initialized with random binary argument strings.
This means that for each argument independently there is a fifty-fifty chance to be \rev{activated} ($a_k = 1$) in the beginning.
Consequently, after projection onto the attitude scale by $C$, the initial distribution of attitudes with respect to the two issues is a binomial distribution with \rev{expected mean zero}.
Depending on the evaluative structure and in particular the evaluative overlap, the initial attitudes on the two issues are already correlated.
As shown in the next section, this initial correlation is enforced in the argument exchange process especially under conditions of polarization.

\rev{
\textbf{Simulation Approach.}
The main aim of this section is to provide an understanding of the kind of processes the model can give rise to. 
For this purpose, we adopt a phenomenological approach that goes deeper into two out of the twelve cases and provide an overview of all cases in the end of the section.
In the two cases where we go into more detail, we first look at the time evolution of the actual opinion distribution of a single realization to provide intuition about the dynamical behavior of the model.
We show that polarization in terms of a bi-modal opinion distribution on the issues and alignment across issues can be a stable outcome of the argument exchange process if homophily is strong, or a transient pattern if homophily is less strong.
Secondly, we compare the opinion distribution after 1000 iterations for different $\beta$.
Notice that the model is implemented such that $N/2$ random pairs are drawn without replacement in single iterations which means that each agent is either sender or receiver during each step.
Especially when population size increases, one can argue that transient opinion landscapes are empirically more relevant than the final absorbing states of the model \citep{Banisch2010empirical}.
Comparing the resulting opinion distributions for single-issue homophily (option 3 above) and the Manhattan distance on both issues (option 1 above) highlights the effect of different distance measures on the specific patterns of polarization and issue alignment.
}

\rev{
\textbf{Model Availability.}
We hence concentrate on specific patterns in the opinion distribution of exemplary model runs that represent what to our experience are typical realizations for the different cases, and provide in this way a mesoscopic picture of model behavior. 
To complement this approach, an online implementation of the model available under \cite{BanischDemos} accompanies the paper\footnote{Direct link: http://www.universecity.de/demos/SCSIssueAlignment.html (Notice that running the model requires a Browser with WebGL support.)}.
In this demonstration, users can interactively explore the effects of different distance measures and bounded confidence thresholds on the argument communication model with two issues.
A wide range cognitive--evaluative maps (far beyond the three cases studied here) can be created by setting the number of congruent and incongruent evaluative connections between arguments and issues.\footnote{For technical reason, enabling a flexible random setup of these maps required that the number of arguments is fixed and with 20 slightly larger compared to the cases studied here.}
While this allows to explore the cases of different argumentative overlap presented here, the online model provides additionally the possibility to include heterogeneity in the cognitive--evaluative maps, and the generation of different evaluative maps for different subpopulations in particular.
}

\subsection{Two Strongly Coupled Issues}
\label{sec:strong}

To illustrate the dynamical behavior of the argument exchange model we first concentrate on a specific combination that highlights the two properties in the distribution of opinions our paper aims to address.
Therefore, we consider the case of two issues that are strongly interrelated in an incongruent way.
The respective evaluative structure is shown on the bottom of Fig. \ref{fig:Settings}.
There are 4 argument dimensions that are relevant for both issues.
Two of those assign a positive weight to the first $c_{k1} = 1$ and a negative weight to the second issue $c_{k2} = -1$ and for the other two it is the other way around.
This means that if one of these facts is believed $a_{sk} = 1$ by an agent $s$ it will have a positive impact regarding its attitude on one and a negative impact on the evaluation of the other issue.
With respect to homophily we assume in this section that the first issue >>polarizes<< meaning that agent with different opinions regarding the first issue are less likely to interact.
The second issue plays no role in that.
The bounded confidence threshold $\beta$ is used to modulate the strength of homophily with respect to the first issue.

\begin{figure}[ht]
 \centering
 \includegraphics[width=0.99\linewidth]{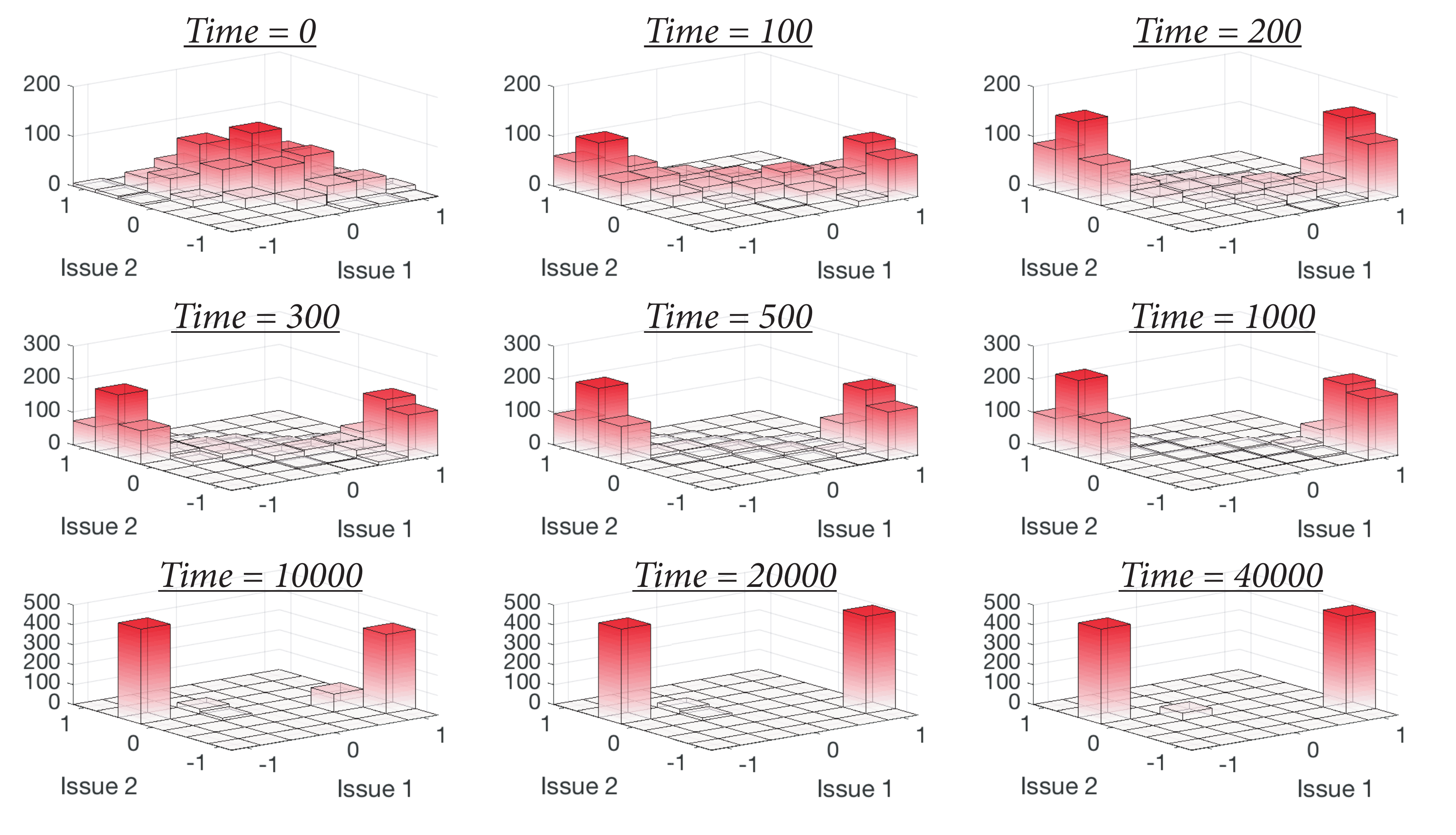}
 \caption{Attitude distribution of 1000 agents in the two-dimensional attitude space during different times of the process. ($\beta = 1$)}
 \label{fig:Series01}
 \end{figure}

Fig. \ref{fig:Series01} shows the distribution of opinions on the two issues at different times of the process starting with the initial distribution on the upper left.
As noted above, the random initialization of arguments leads to a binomial initial distribution of opinions for each of the issues.
The extremes of the opinion spectrum are only rarely populated.
The strong incongruent overlap encoded in the evaluative structure in this setting already induces a negative correlation between the initial opinions on both issues.
Already after 100 steps this correlation pattern becomes considerably more pronounced, the spread of the distribution increases and the extremes become populated.
Note that after this relatively short time the most extreme opinions on issue 1 (homophily-relevant) are adopted by the majority of agents but that also the second issue is polarized, albeit not that strongly.
This strong pattern of bi-polarization is accentuated during subsequent steps of the simulation and agents with intermediate attitudes become very rare.
The strong negative coupling between the issues constrains the two different opinion groups to specific combinations of attitudes.
The two groups develop opposing views not only with respect to the >>polarizing<< first issue but also with respect to the second one.

An example for such a setting might be two policy proposals that are discussed as competing alternative solution to the same problem.
If one believes in coal as the future technology for electric power production and supports public investments into that area one probably disfavors subsidies in renewable energy production technology.
Of course, our model provides only very stylized representations of such complex issues but it reveals a possible logic behind the formation of certain constellations of attitudes.
It also shows how two groups that develop opposing attitudes tend to selectively believe and adopt facts that support their respective views if their interaction behavior is guided by their current attitudinal stance deciding about interlocutors whose arguments can be taken seriously.

\begin{figure}[ht]
 \centering
 \includegraphics[width=0.99\linewidth]{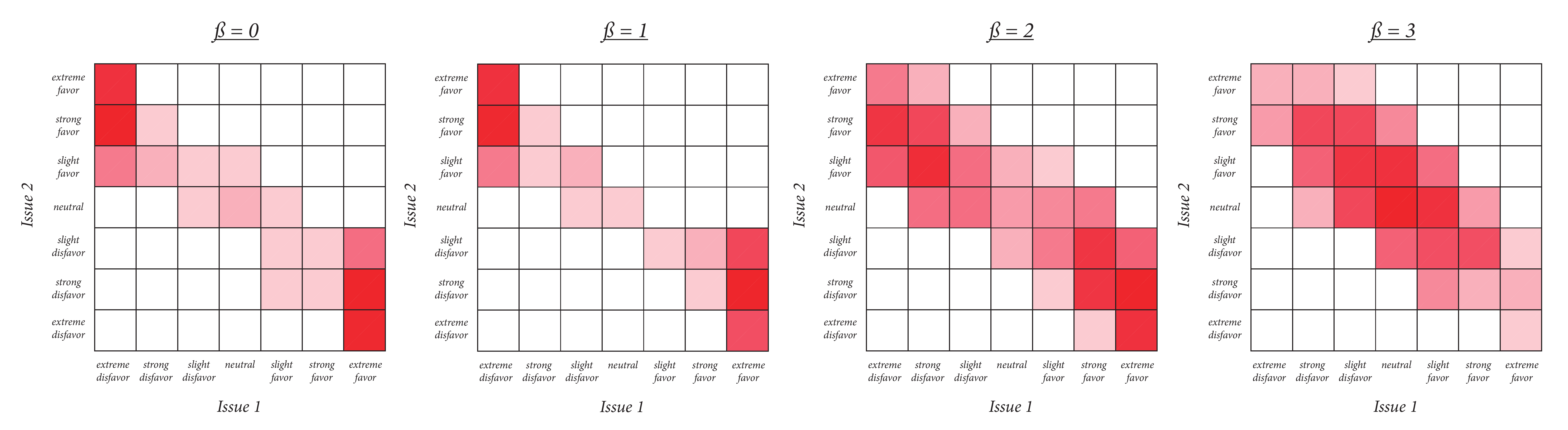}
 \caption{Attitude distribution of 1000 agents after 1000 iterations in the two-dimensional attitude space for different values of the bounded confidence threshold ($\beta = 0, 1, 2, 3$). The pattern of ideological coherent polarization is stable for a wide set of $\beta$.}
 \label{fig:BCSeries02}
 \end{figure}

In our model, the bounded confidence threshold $\beta$ decides about the strength of this social interaction bias.
For the given setting, Fig. \ref{fig:BCSeries02} shows the opinion distributions after 1000 iterations for $\beta$ ranging from zero to three (left to right).
Notice that $\beta = 0$ indicates that only agents with exactly the same opinion on issue 1 engage into the argument exchange process.
Likewise, for $\beta = 2$ agents that are differ at most 2 points on the 7 item attitude scale from extreme disfavor to extreme favor interact.
This means, for instance, that a neutral agent will interact with strong supporters but not with extreme ones and that agents that weakly favor one option will still interact with agents who weakly disfavor it.

As Fig. \ref{fig:BCSeries02} shows, a polarization pattern emerges up to the case of $\beta = 2$.
\rev{
This effect is stronger on the salient first issue, but due to the strongly incongruent evaluative structure also rather pronounced on the second one.
The population divides into two opposing groups of agents with a supportive stance regarding one and a negative stance regarding the other issue. 
This is a strong form of opinion alignment.
}
If the threshold increases, homophily is not strong enough to lead to opinion bi-polarization \rev{and the population approaches a moderate consensus in the long run}.
\rev{
Notice that the argument exchange process gives rise to only two outcomes in this mode of homophily: bi-polarization if $\beta$ is small and consensus if $\beta$ is large.
Noteworthy, a bi-polarized organization of opinions is observed even for $\beta = 0$.
This is in contrast to other models of bounded confidence \citep{Hegselmann2002opinion,Deffuant2000mixing,Lorenz2007continuous2} with a typical transition from complete fragmentation for small confidence thresholds to polarization for intermediate ones and consensus if $\beta$ is large.
}

\subsection{Two Weakly Congruent Issues}

\begin{figure}[ht]
 \centering
 \includegraphics[width=0.89\linewidth]{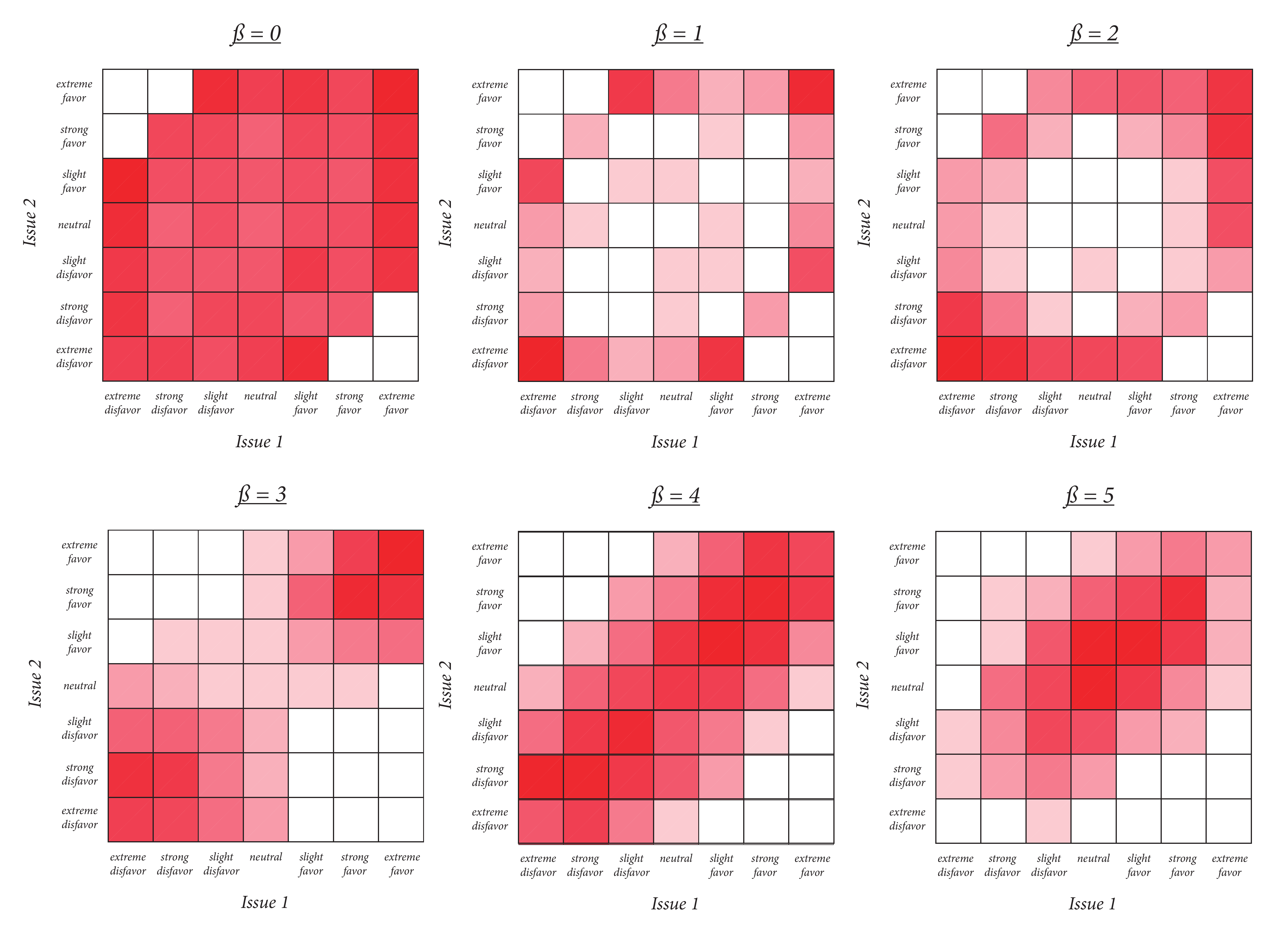}
 \caption{Attitude distribution of 1000 agents after 1000 iterations in the two-dimensional attitude space for different values of the bounded confidence threshold ($\beta = 0, 1, 2, 3,4,5$). The Manhattan distance is used in the homophily mechanism.}
 \label{fig:BCSeries03}
 \end{figure}

Let us consider another example slightly more carefully and look at the case of two weakly coupled issues as shown in the middle in Fig. \ref{fig:Settings}.
The overlap in the evaluative structure indicates that a slightly positive relation between the opinions will result from such interdependencies.
In terms of homophily, we consider now the Manhattan distance in the two-dimensional opinion space and assume that argument exchange takes place only if the $|o_s(1)-o_r(1)| + |o_s(2)-o_r(2)| < \beta$.
The opinion distribution after 1000 iterations is shown in Fig. \ref{fig:BCSeries03} for $\beta$ ranging from zero to five.

We observe two interesting transitions in this case.
First, if $\beta$ is small (here if $\beta \leq 2$), distributions emerge in which the neutral positions on the two issues are more and more sparsely occupied.
\rev{
This is compatible with the analysis of the 2D bounded confidence model by \cite{Lorenz2003mehrdimensionale}.
}
Especially the case of $\beta = 1$ shows that there may be several highly populated configurations of opinions arranged elliptically around the center.
\rev{
As opposed to the previous case of homophily with respect to a single issue, a more fragmented opinion profile is observed for small $\beta$ if the two issues are taken into consideration for homophily.
}
This means that despite the positive coupling of the two issues, groups of agents may adopt a negative stance on one and a positive opinion on the other issue.
As shown in the last row of Fig. \ref{fig:OverviewTable} the shape of this depends on the strength of the evaluative overlap between the two issues.
  
As the bounded confidence threshold increases beyond $\beta = 2$ a different pattern emerges which is more closely related to a typical bi-polarized opinion distribution.
Two groups emerge that develop consistently opposing standpoints on the two issues with negative or positive opinions regarding both issues.
Notice that for $\beta = 4$ the distribution along this negative/negative to positive/positive dimension is much more flat and intermediate neutral opinions are still present after 1000 iterations.
\rev{
While a polarized state is stable for $\beta = 2$ it becomes a transient phenomena for $\beta = 4$.
}

\begin{figure}[ht]
 \centering
 \includegraphics[width=0.99\linewidth]{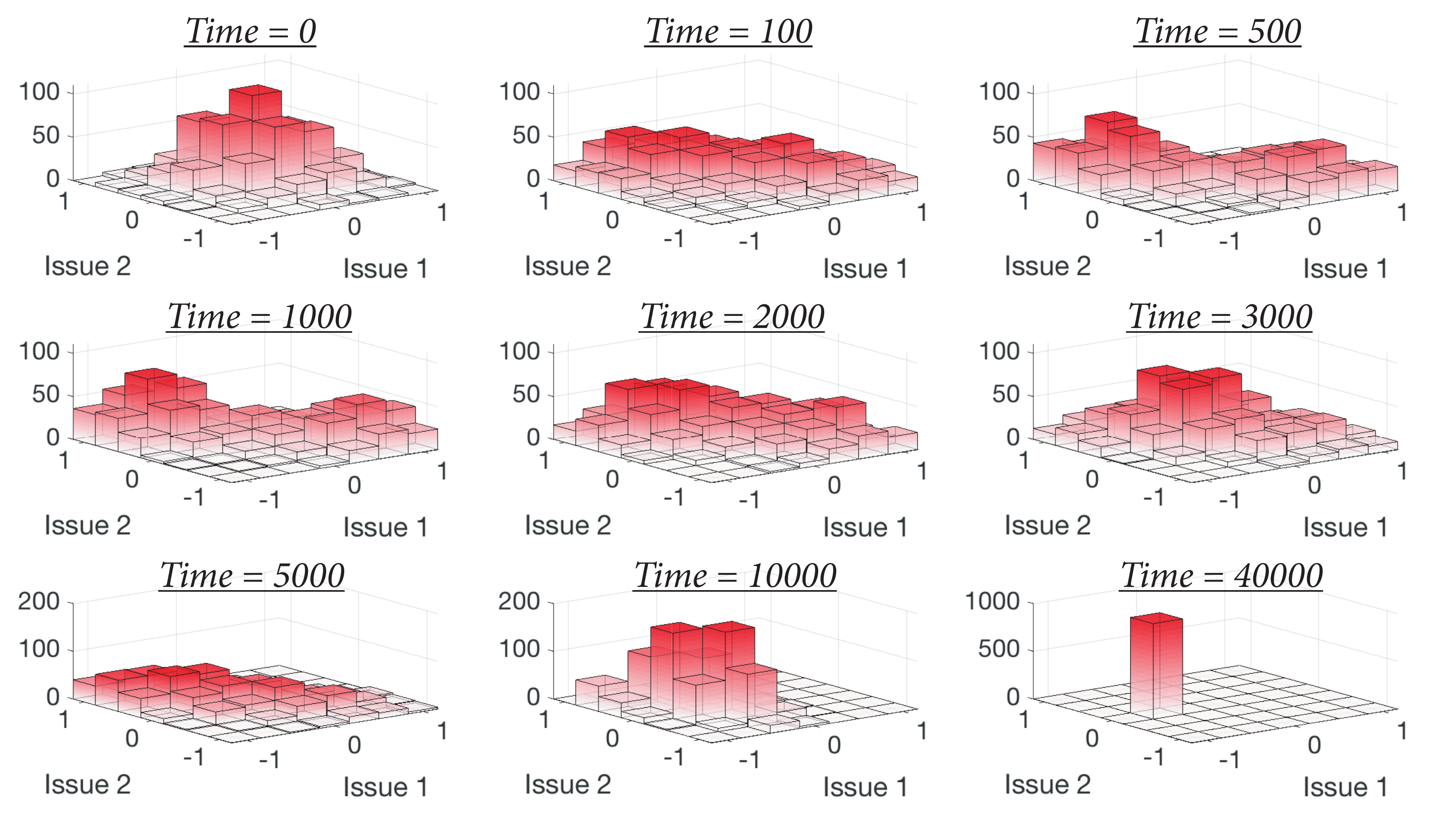}
 \caption{Time evolution of the attitude distribution for $\beta = 4$. The persistence of intermediate opinions resolves the short-term polarization in the long run.}
 \label{fig:SeriesBeta4}
 \end{figure} 

This is shown in Fig. \ref{fig:SeriesBeta4} where the time evolution of the opinion distribution is considered.
Although a clear pattern of bi-polarization is visible after 500 steps, intermediate moderate opinions can be sustained in this parameter setting.
These intermediate agents help to maintain a  >>flow of arguments<< between the two groups which circumvent the emergence of isolated argument pools that would lead to persistent inter-group polarization.  
As a result of this persistent exposure to diverse arguments, agents with extreme opinions become more moderate again.
\rev{
In fact, the similarity between the initial opinion distribution and the distribution after 3000 steps is remarkable.
The fact that very similar attitude distributions can lead to increasing polarization in one case and to moderation and long--term consensus in the other under argument communication indicates that a very interesting reorganization of beliefs has taken place in this initial phase of increasing polarization.
It also highlights that the cognitive--evaluative layer -- beliefs, evaluations, arguments, etc. -- underlying patterns of public opinion may be of crucial importance for a better understanding of opinion change.
}

\subsection{Summary of the Model Behavior}
\label{sec:resultssummary}

\begin{figure}[h]
  \centering
  \includegraphics[width=0.99\linewidth]{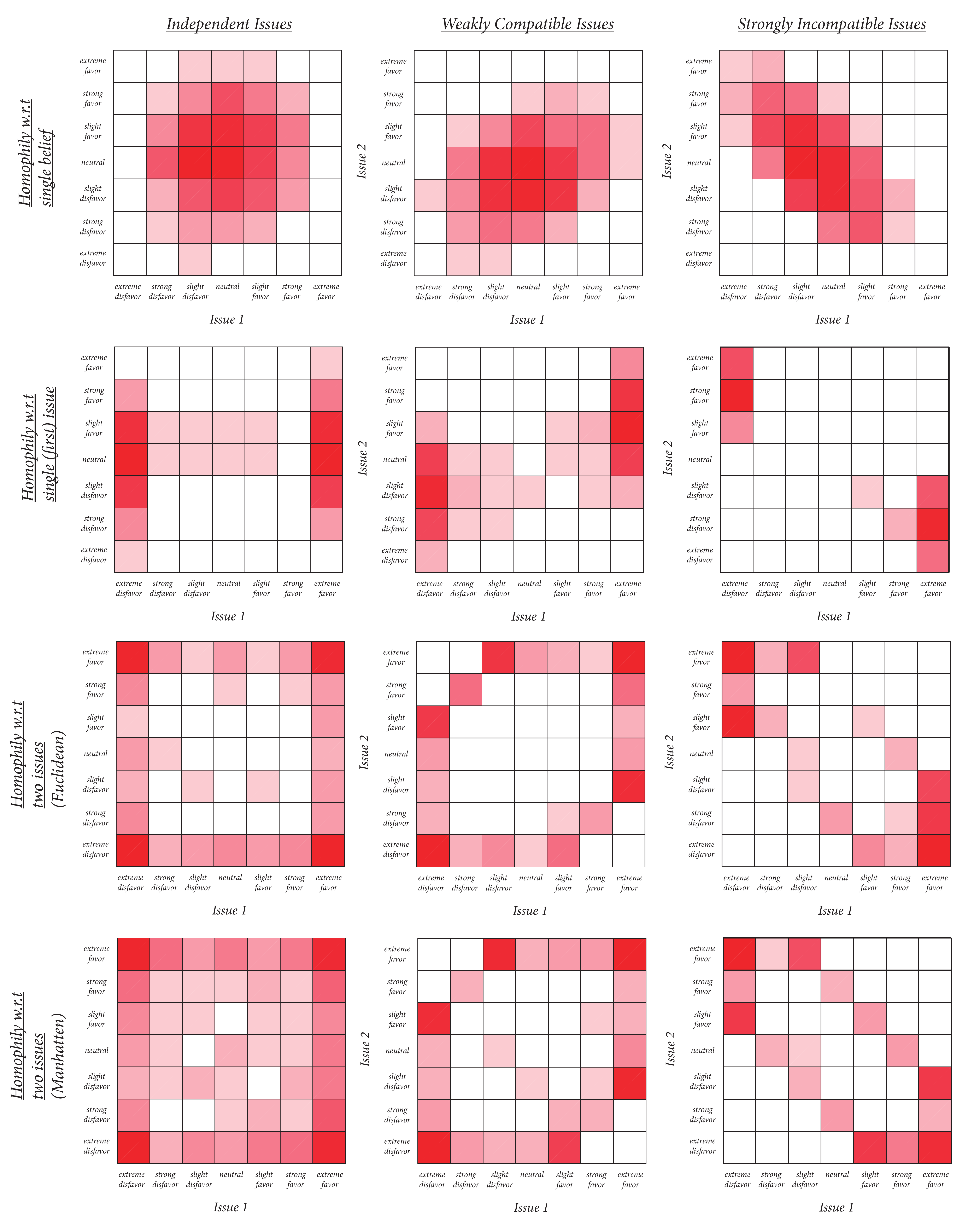}
  \caption{Overview over the model behavior in the different conditions described in Sec. \ref{sec:settings}. The distribution of $N = 1000$ after 1000 iterations is shown with $\beta = 1$.}
  \label{fig:OverviewTable}
  \end{figure}

In Fig. \ref{fig:OverviewTable}, an overview of the model behavior in the different settings described in Sec. \ref{sec:settings} is provided.
Here we show the distribution of opinions after 1000 iteration for a relatively small $\beta = 1$.
The three columns of this figure correspond to the three different evaluative maps shown in Fig. \ref{fig:Settings}.
In the first row the two issues are completely independent, in the second column the issues are weakly interrelated as in the previous section, and the third column represents the strongly coupled case considered in Sec. \ref{sec:strong}.

The rows correspond to four different ways to take into account opinion homophily.
In the first row, we consider the case that two agent engage in the argument exchange process only if their belief regarding a single fact ($k=5$ in this case) is equal.
The idea behind this is that a single belief may signal an unacceptable stance of the interlocutor (such as, for instance, neglecting that CO2 emissions cause global warming).
However, under the argument exchange dynamics incorporated into our model, this is generally not sufficient for polarization to emerge.
The second row considers the homophily mechanisms studied in Sec. \ref{sec:strong}. 
As expected, a strong pattern of bi-polarization with respect to the >>polarizing<< issue emerges.
The strength of interrelatedness of the two issues encoded in the cognitive-evaluative map governs to what extent the other issues (issue 2 in this case) polarizes >>along with<< the first issue.
\rev{
Noteworthy, in this case the bi-polarized outcome as opposed to fragmentation in other bounded confidence models \citep{Lorenz2007continuous2} is observed even if $\beta$ is very small.
}
The last two rows correspond to two different distance measures that take into account the agents' positions on both issues.
In the third row the Euclidean distance is used and in the forth one the Manhattan distance.
The figure shows that the behavior in terms of the distribution of opinions is very similar.
\rev{
Opinions are arranged around the center with strength and direction of argumentative overlap governing the shape of this pattern.
Notice that a similar effect of a circular pattern around an unpopulated center has been observed in multi--dimensional continuous opinion dynamics with bounded confidence \citep[53]{Lorenz2003mehrdimensionale}.
When issues are interdependent} 
the arrangement of admissible opinions around the center shows that certain opinion configurations are impossible due to the evaluative structure.
Namely, under weak congruent coupling an extremely positive stance on one issues implies that an extremely negative stance with respect to the other is not admissible.
These constraints imposed by the evaluative map are even more substantial under strong coupling where the upper right and the lower left corner of the opinion space cannot be occupied.

The main purpose of this section has been to highlight two essential properties of the argument exchange model when extended to multiple issues.
For this purpose, we have concentrated on a set of stylized settings.
We have shown that the bi-polarizing dynamics of the ACTB \citep{Maes2013differentiation} is recovered by our version.
By showing that the polarization with respect to one issue may force agents to take a specific viewpoint on another issue if they are cognitively related we provide a possible explanation of a further important aspect of opinion polarization: the alignment of opinions across different issues within the two opposing groups of agents.
This aspect of polarization -- that is, an increasing >>ideological uniformity<< -- is at the core of recent empirical studies on political polarization in American public opinion \citep{Dimock2014political}.
\rev{
The comparison of modes of attitude homophily that take into account only one or respectively the two issues suggests that a single particularly >>loaded<< issue may be an important driver towards more pronounced bi-polarization into two ideological aligned camps.
}

\section{An Example with Three Issues}
\label{sec:threeissueexample}

One of the objectives of the model presented throughout this paper is to work towards a framework that allows to connect opinion dynamics modeling to real data on political statements and opinions.
The evaluative structure is compatible with structural expectancy-value models of attitudes \citep{Fishbein1962ab,Fishbein1963investigation,Ajzen2001nature} that are still widely used in survey-based attitude research.
We will exploit this potential in the future but shall conclude this paper with a stylized example of some empirical plausibility.

\begin{figure}[ht]
 \centering
 \includegraphics[width=0.99\linewidth]{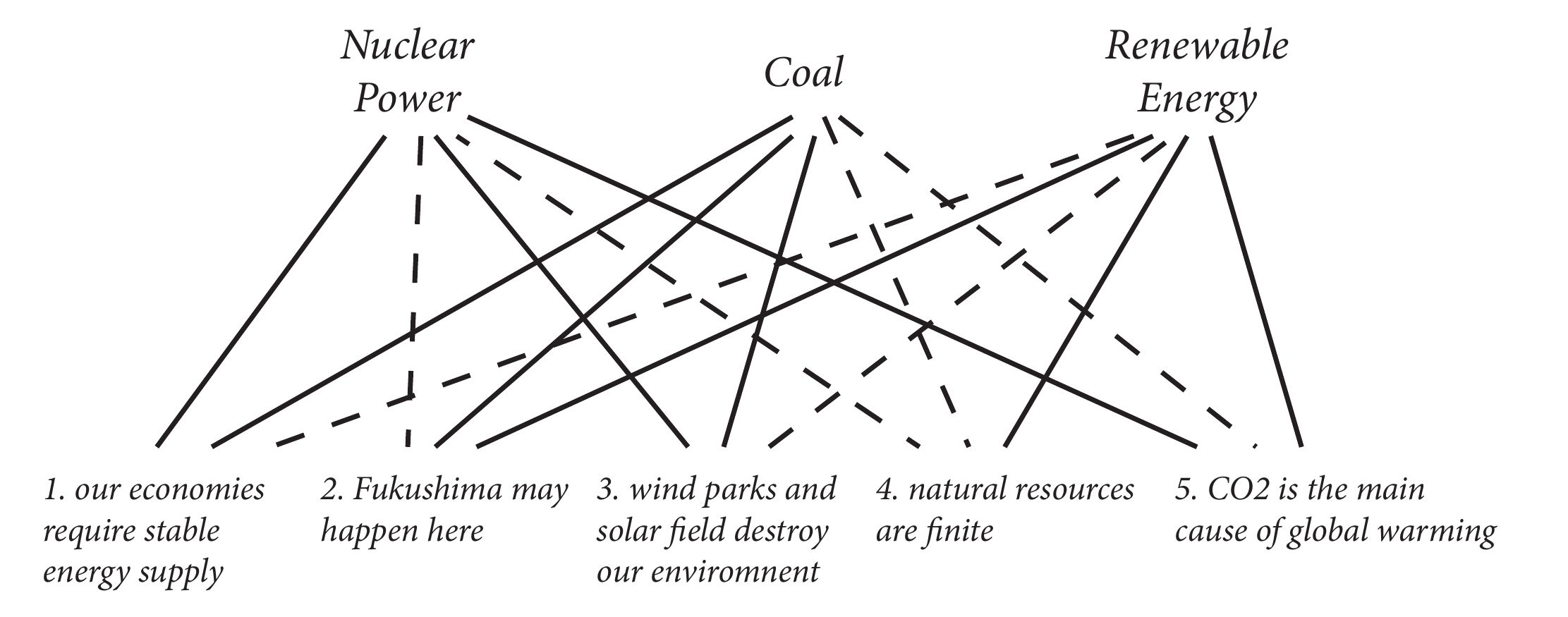}
 \caption{An example with five factual dimensions and three issues. The structure is highly constrained as all arguments are negatively (dashed) or positively (solid) related to all the three issues.}
 \label{fig:SpecificExample01}
 \end{figure}

We consider a population of agents that debates on three different technologies for electric power production: nuclear power ($i_1$), coal ($i_2$) and technologies based on renewable sources ($i_3$).
We allow only five arguments that are related by different degrees to economic stability ($k_1, k_4$), security ($k_2$) and sustainability ($k_3,k_4,k_5$):
\begin{itemize}
\item[$k_1$]
>>The growing economies of the world need reliable and scalable energy sources.<< 
\item[$k_2$]
>>An accident like in Fukushima may happen here.<<
\item[$k_3$]
>>Large wind parks and vast solar fields impair the appearance of our environment.<<
\item[$k_4$]
>>Natural resources are finite.<<
\item[$k_5$]
>>Human-made CO2 emissions are the main cause of global warming.<<
\end{itemize}
The choice of some of these statements has been inspired by the debates on the web page ProCon.org.\footnote{https://alternativeenergy.procon.org in particular.}
As shown in Fig. \ref{fig:SpecificExample01}, all the arguments are linked to all the issues as they are most often used as pro- or con-arguments for one technology against others.
Notice that the arguments have been chosen with some care so that each issue has two negative and three positive connections.

\begin{figure}[ht]
 \centering
 \includegraphics[width=0.99\linewidth]{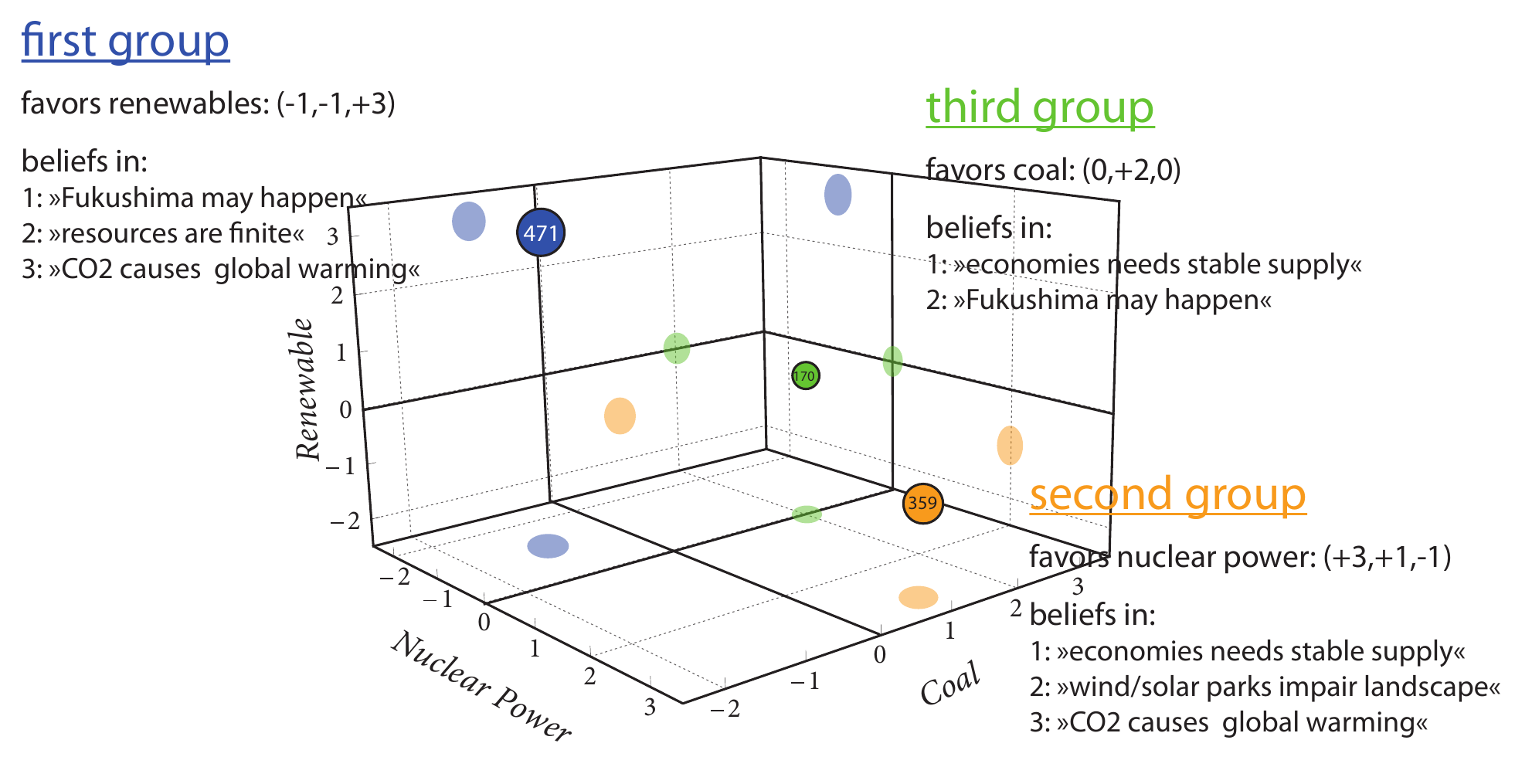}
 \caption{Result of a realization of the model with the evaluative structure shown in Fig. \ref{fig:SpecificExample01}. Three groups emerge in the argument exchange process. One (blue, 471 agents) strongly supports renewable energy opposing coal and nuclear power. A second one (yellow, 359 agents) strongly supports nuclear power but is also slightly positive with respect to coal. A third group (green, 170 agents) favors coal and is neutral with respect to nuclear power and renewables.}
 \label{fig:Figure01}
 \end{figure}

We run this model using the Manhattan distance in the homophily condition and with $\beta = 3$.
Fig. \ref{fig:Figure01} shows the opinion configurations after 20000 steps.
In this case, the population has converged and no further argument exchange is possible due to the bounded confidence threshold.
In this simulation three different groups emerge.
The first one with almost one half of the population (471 agents) develops an opinion that is extremely supportive of renewable energy technology (+3) and has a negative opinion regarding the other two (-1).
The string of beliefs of this group is $(0 1 0 1 1)$, meaning that they belief that nuclear accidents may happen again (2nd argument), that resources are finite (4th argument) and that CO2 causes global warming (5th argument).
That is, the process of repeated argument exchange has led this group to believe only in those three arguments that are positively related to renewable energy.
The second group of 359 agents strongly support nuclear power (+3) while downgrading renewables (-1) but not coal (+1).
The associated belief string of this subpopulation is $(1 0 1 0 1)$.
As for the first group, the interaction process has led to selective beliefs in arguments supporting nuclear power while neglecting the second and fourth argument concerning nuclear accidents and finite resources.
A third smaller group of agents (170) supports coal (+2) while being neutral (0) with respect to the other two technologies.
The beliefs in this group have settled on ($1 1 0 0 0$), that is, the first argument related to stable supply and the second one related to nuclear accidents.

Although we will discuss implications and limitations of the model in the next section, a short comment on the significance of this example shall be made here. 
We do certainly not want to claim that this restricted argument set is suited to accurately represent the real debate around those issues.
But the example is informative about how our argument exchange model can be applied and what kind of results we can expect if real argument data is fed into it.
Very often the intertwining of different arguments about different issues is rather hard to disentangle and its consequences for deliberative communication are unclear.
The proposed model can improve our understanding how underlying structures (of beliefs, of argumentation, of evaluative connotations) affect debates like the one caricatured here.

For instance, further simulations have revealed that the emergence of a blue group (strong renewable supporters) is very robust and that the distance in attitude space between this group and the supporters of the two other technologies is generally larger than the distance between these other two.
\rev{
In the example sketched here this primary dimension of polarization between renewable supporters and opposers is very stable.
Also an approximately fifty-fifty division of the population along this axis is a robust feature.
}
The reason for this is an asymmetry in the evaluative structure (Fig. \ref{fig:SpecificExample01}) in which the argument 4 (>>natural resources are finite<<) is the only one which contributes negatively to two of the issues (namely, coal and nuclear power).
In such circumstances, an alliance between the supporters of coal and nuclear power (e.g. to gain majority) becomes considerably more likely than the other coalitions.
\rev{
This is also consistent with the model outcomes for larger $\beta$ (i.e. $\beta = 4$) where in virtually all cases two groups of similar size emerge that resemble such a formation of coalitions.
}

\section{Discussion}
\label{sec:discussion}

With this paper, we contribute to a relatively new development to introduce a cognitive level into models of opinion dynamics.
While most previous models operate on rather abstract opinion spaces ranging from binary states \citep{Sznajd-Weron2000opinion,Galam2005local,Banisch2014microscopic} to multi-dimensional \citep{Axelrod1997dissemination,Baldassarri2007dynamics,Banisch2010acs} and continuous state spaces \citep{Deffuant2000mixing,Hegselmann2002opinion}\footnote{See \citep{Castellano2009statistical,Flache2017models} for more encompassing overviews.}, the incorporation of belief systems \citep{Friedkin2016network,Parsegov2016novel}, structural representations of attitudes \citep{Urbig2005dynamics,Maes2013differentiation} and cognitive networks \citep{VanOverwalle2006talking,Wolf2012adoption,Wolf2015changing} is a relatively new development \citep[see also][]{Conte2014agent}.
By allowing to integrate sometimes very specific information \citep[see esp. ][]{Wolf2015changing} on the issues addressed by the model, this is a promising development to answer the fundamental critique that opinion dynamics lacks links to real data \citep{Sobkowicz2009modelling}, see also \cite{Flache2017models}.
\rev{
In this paper, we contribute to this development using the model by \cite{Maes2013differentiation} as a starting point.
The multilevel opinion structure assumed in the original paper is extended to multiple cognitively related issues by assuming a cognitive-evaluative map which models the evaluative meaning of different arguments with respect to the different issues.
This conceptualization of attitudes is closely related to structural theories of attitudes \citep{Ajzen2001nature}, generally compatible with standard survey techniques and conceptualizes attitudes and arguments in a way closely related to recent experimental designs to measure the influence of arguments \citep{Kobayashi2016relational,Shamon2019arguments}.
It therefore provides a promising framework to establish the connection between models and empirical cases.
Moreover, it is also compatible with the above-mentioned connectionist models that employ cognitive networks \citep[273]{VanOverwalle2005connectionist} and therefore an optimal trade-off between model complexity and parsimony, rich enough to incorporate relevant aspects of the conceptual structure underlying a real debate.}

We illustrate this potential by the example studied in the last section, but the main focus of this paper has been to understand the basic properties of such a model.
The model is already complex and involves quite a number of parameters and design choices, most importantly: the parameterization of the evaluative structure, the conceptual overlap and different forms of homophily at the attitude level.
It is important to understand the impact of these modeling choices before the model can be used in more applied scenarios.
We have followed a computational approach here by systematically analyzing 12 combinations of 4 homophily modes and 3 paradigmatic evaluative structures, but also analytical strategies developed in theoretical biology can be applied to simple cases \citep{Banisch2018math}.

Most sociologist agree that one should include psychological complexity into models addressing social aggregations and collective phenomena only to the extent that they substantially contribute to the explanation of the phenomenon at stake \citep{Lindenberg1992method,Kron2004general,Kroneberg2005definition}.
On the one hand side, we follow this tradition by developing a very simple model of argument communication that simplifies the original model of ACTB by \cite{Maes2013differentiation} in two regards.
Namely, we assume a process of argument exchange in which beliefs are directly transmitted from a sender to a receiver (>>copied<<) and we simplify the homophily mechanism by relying on the concept of bounded confidence.
While the psychological precision of these assumptions at the inter-individual level can be disputed \citep[and sometimes is, ][]{Mueller2018cognitive} our results show that, at the aggregate level, the essential dynamical properties of ACTB are preserved under the parsimonious design choices made here.
\rev{
Notice also that in finite size systems occasional deviations from the sharp bounded confidence scheme may lead to homogeneity in the long run \citep{Maes2010individualization}.
}
In another regard, we take a step towards a more complex model by focusing on different cognitive-evaluative maps that link beliefs to attitudes.
This generates a considerable degree of freedom in our model and, in fact, would lead to a combinatorial explosion of the number of free parameters if these structures were purely individual.

However, in our case we consider that these evaluation structures have been acquired in the same socio-cultural context and are collectively shared by all agents.
While from the methodological point of view this simply reduces the number of parameters and makes systematic computational analysis possible, there is also a tradition in Sociology to view culture as shared meaning structures \citep{Berger1970gesellschaftliche,Schuetz2017strukturen} and their integration in form of cognitive networks seems a viable approach \citep[cf. ][359, and see also \cite{Strauss1997cognitive}]{Kroneberg2005definition}.
Understood in this way, as >>ideal-typical<< 
cognitive representations of evaluative meaning acquired in long processes of socialization, these structures become very relevant to the analysis of public opinion formation and are, in our model, essential for the explanation of issue alignment.
Furthermore, the model entails the possibility to consider different sub-population or sub-cultures with differing cognitive maps and is therefore suited to explore the impact of cultural differences (operationalized in this way) on deliberative argument exchange process \citep[a first attempt is made in][]{Banisch2018math}.
This also allows to link the model to recent empirical work on the identification of different belief systems within different social strata or sub-cultures \citep[][see also \cite{Converse1964nature}]{Baldassarri2014ideologies, Daenekindt2017how} and the qualitative differences with respect to issue alignment in particular \citep[cf.][Figure 7]{Goldberg2011mapping}. 
\rev{
The online implementation of the model \citep{BanischDemos} entails the possibility to define subgroups with different evaluative structures.
}

\rev{Models are caricatures of real social processes and for opinion dynamics, where empirical measurement is in itself a hard task, this is even more evident.
The model put forth here concentrates on a process of exchange of arguments and leaves many important things out.
Identity, sometimes argued to play a pivotal role in opinion making \citep{Achen2017democracy} is not included.
We have considered the case of a single argument that might signal identity, but a more appropriate incorporation of identity certainly requires the integration of different mechanisms to account for the complex interaction of attitude and identity.
Well-established psychological effects relating motivated reasoning and biased argument processing to attitude polarization \citep{Lord1979biased,Kunda1990case,Taber2009motivated,Kobayashi2016relational} are missing as well.
On the other hand, the homophily mechanism by which agents avoid exchange with others that think differently can also be interpreted as a tendency to judge their arguments as not reliable or relevant giving favor to arguments that are coherent with the own attitude.
Social structure beyond opinion homophily is also left out of the model.
We have done some simulation experiments that include different social interaction structures and found no indication that it changes the results reported here, mainly due to the fact that very similar processes of opinion alignment and polarization are observed within different communities. 
Yet it might be very interesting to reconsider the effects of demographic crisscrossing addressed in \cite{Maes2013short} when >>demographic faultlines<< (716) mark differences in the evaluative maps such that arguments are interpreted differently.}

\rev{
Given that many different things could be -- and have been - included into models of opinion dynamics, it is \emph{a fortiori} important to devise scenarios in which models can be tested and different assumptions confronted.
This is very difficult with most existing models as opinions are usually >>void of meaning<< and no correspondence to the thematic dimensions of real debates is sought.
The model developed in this paper points out a possible direction to overcome this deficiency by mapping arguments used in real debates and the opinions they support.
This allows to embed the model in \emph{empirically informed scenarios} so that they can be validated on specific cases of discourse that involve opinions.
The energy technology example sketched in the previous section provides an illustration of such a setting.
Using survey data or argument data extracted from text with precision language processing techniques \citep{Steels2017basics,VanEecke2018exploring} to inform the underlying opinion structure will open up completely new ways of model validation.
The argument exchange mechanism studied here might be appropriate for some thematic complexes, others might teach us that different aspects such as group identity or morality have to be taken more seriously. 
In any case, research on opinion dynamics will greatly benefit from such a program.
}

\section{Conclusion}
\label{sec:conclusion}

This paper presents a model of argument exchange dynamics that extends the >>argument communication theory of bi-polarization (ACTB)<< proposed in \cite[2]{Maes2013differentiation}.
Our main contribution is to show that an argument exchange account of social influence dynamics can provide a useful framework for modeling processes of issue alignment \citep{Baldassarri2014ideologies,Dimock2014political} by which attitudes on a set of issues become correlated along (ideological) dimensions.
While the main focus in recent model-based studies of polarization \cite[e.g.][]{Dandekar2013biased,Friedkin2015problem,Maes2015will,Duggins2017psycologically,Banisch2017opinion} has been to explain the emergence of a bi-polar opinion distribution on a single issue, empirically motivated studies of polarization \citep[e.g][]{DiMaggio1996have,Baldassarri2014ideologies,Dimock2014political} indicate that the alignment of attitudes across several issues is at least an equally important signature of polarization.
To our knowledge this is the first modeling account that addresses these inter-issue dependencies and constraints in an explicit way.

\section*{Acknowledgment}

This project has received funding from the European Union’s Horizon 2020 research and innovation programme under grant agreement No 732942 (Opinion Dynamics and Cultural Conflict in European Spaces -- www.\textsc{Odycceus}.eu).

\section*{References}


\small

\begin{thebibliography}{84}
\providecommand{\natexlab}[1]{#1}
\providecommand{\url}[1]{\texttt{#1}}
\expandafter\ifx\csname urlstyle\endcsname\relax
  \providecommand{\doi}[1]{doi: #1}\else
  \providecommand{\doi}{doi: \begingroup \urlstyle{rm}\Url}\fi

\bibitem[Achen and Bartels(2017)]{Achen2017democracy}
C.~H. Achen and L.~M. Bartels.
\newblock \emph{Democracy for realists: Why elections do not produce responsive
  government}, volume~4.
\newblock Princeton University Press, 2017.

\bibitem[Ajzen(2001)]{Ajzen2001nature}
I.~Ajzen.
\newblock Nature and operation of attitudes.
\newblock \emph{Annual Review of Psychology}, 52\penalty0 (1):\penalty0 27--58,
  2001.

\bibitem[Axelrod(1997)]{Axelrod1997dissemination}
R.~Axelrod.
\newblock The dissemination of culture: A model with local convergence and
  global polarization.
\newblock \emph{The Journal of Conflict Resolution}, 41\penalty0 (2):\penalty0
  203--226, 1997.

\bibitem[Bacharach and Gambetta(2001)]{Bacharach2001trustsigns}
M.~Bacharach and D.~Gambetta.
\newblock Trust in signs.
\newblock In \emph{Trust in Society}, pages 148--184. Russell Sage Foundation,
  2001.

\bibitem[Bakshy et~al.(2015)Bakshy, Messing, and Adamic]{Bakshy2015exposure}
E.~Bakshy, S.~Messing, and L.~A. Adamic.
\newblock Exposure to ideologically diverse news and opinion on facebook.
\newblock \emph{Science}, 348\penalty0 (6239):\penalty0 1130--1132, 2015.


\bibitem[Baldassarri and Bearman(2007)]{Baldassarri2007dynamics}
D.~Baldassarri and P.~Bearman.
\newblock Dynamics of political polarization.
\newblock \emph{American Sociological Review}, 72\penalty0 (5):\penalty0
  784--811, 2007.

\bibitem[Baldassarri and Goldberg(2014)]{Baldassarri2014ideologies}
D.~Baldassarri and A.~Goldberg.
\newblock Neither ideologues nor agnostics: Alternative voters’ belief system
  in an age of partisan politics.
\newblock \emph{American Journal of Sociology}, 120\penalty0 (1):\penalty0
  45--95, 2014.

\bibitem[Banisch(2014)]{Banisch2014microscopic}
S.~Banisch.
\newblock From microscopic heterogeneity to macroscopic complexity in the
  contrarian voter model.
\newblock \emph{Advances in Complex Systems}, 17\penalty0 (05):\penalty0
  1450025, 2014.

\bibitem[Banisch(2019)]{BanischDemos}
S.~Banisch.
\newblock {www.UniVerseCity.de -- Demo Section: Interactive Exploration of
  Opinion Models}.
\newblock http://www.universecity.de/index.php?site=demos, March 2019.

\bibitem[Banisch and Ara{\'u}jo(2010)]{Banisch2010empirical}
S.~Banisch and T.~Ara{\'u}jo.
\newblock On the empirical relevance of the transient in opinion models.
\newblock \emph{Physics Letters A}, 374\penalty0 (31):\penalty0 3197--3200,
  2010.

\bibitem[Banisch and Olbrich(2019)]{Banisch2017opinion}
S.~Banisch and E.~Olbrich.
\newblock Opinion polarization by learning from social feedback.
\newblock \emph{The Journal of Mathematical Sociology}, 43(2), 2019.

\bibitem[Banisch et~al.(2010)]{Banisch2010acs}
S.~Banisch, T.~Araujo, and J.~Louçã.
\newblock Opinion dynamics and communication networks.
\newblock \emph{Advances in Complex Systems}, 13\penalty0 (1):\penalty0
  95--111, 2010.

\bibitem[Banisch et~al.(forthcoming)Banisch, Tran, and
  Olbrich]{Banisch2018math}
S.~Banisch, T.~D. Tran, and E.~Olbrich.
\newblock Argument exchange dynamics in a population with divergent mindsets.
\newblock forthcoming.
\newblock A first version of this paper has been presented at the Conference on
  Complex Systems 2018, Thessaloniki.

\bibitem[Berger and Luckmann(1970)]{Berger1970gesellschaftliche}
P.~L. Berger and T.~Luckmann.
\newblock \emph{Die gesellschaftliche Konstruktion der Wirklichkeit. Eine
  Theorie der Wissenssoziologie.}
\newblock Frankfurt a. M., 1970.

\bibitem[Bramson et~al.(2016)Bramson, Grim, Singer, Fisher, Berger, Sack, and
  Flocken]{Bramson2016disambiguation}
A.~Bramson, P.~Grim, D.~J. Singer, S.~Fisher, W.~Berger, G.~Sack, and
  C.~Flocken.
\newblock Disambiguation of social polarization concepts and measures.
\newblock \emph{The Journal of Mathematical Sociology}, 40\penalty0
  (2):\penalty0 80--111, 2016.

\bibitem[Burnstein and Vinokur(1975)]{Burnstein1975person}
E.~Burnstein and A.~D. Vinokur.
\newblock What a person thinks upon learning he has chosen differently from
  others: Nice evidence for the persuasive-arguments explanation of choice
  shifts.
\newblock 1975.

\bibitem[Burnstein and Vinokur(1977)]{Burnstein1977persuasive}
E.~Burnstein and A.~D. Vinokur.
\newblock Persuasive argumentation and social comparison as determinants of
  attitude polarization.
\newblock 1977.

\bibitem[Byrne(1961)]{Byrne1961interpersonal}
D.~Byrne.
\newblock Interpersonal attraction and attitude similarity.
\newblock \emph{The Journal of Abnormal and Social Psychology}, 62\penalty0
  (3):\penalty0 713, 1961.

\bibitem[Castellano et~al.(2009)Castellano, Fortunato, and
  Loreto]{Castellano2009statistical}
C.~Castellano, S.~Fortunato, and V.~Loreto.
\newblock Statistical physics of social dynamics.
\newblock \emph{Reviews of Modern Physics}, 81\penalty0 (2):\penalty0 591,
  2009.

\bibitem[Conte and Paolucci(2014)]{Conte2014agent}
R.~Conte and M.~Paolucci.
\newblock On agent-based modeling and computational social science.
\newblock \emph{Frontiers in Psychology}, 5:\penalty0 668, 2014.

\bibitem[Converse(1964)]{Converse1964nature}
P.~E. Converse.
\newblock The nature of belief systems in mass publics.
\newblock \emph{Critical Review}, 18\penalty0 (1-3):\penalty0 1--74, 1964.


\bibitem[Daenekindt et~al.(2017)Daenekindt, de~Koster, and van~der
  Waal]{Daenekindt2017how}
S.~Daenekindt, W.~de~Koster, and J.~van~der Waal.
\newblock How people organise cultural attitudes: cultural belief systems and
  the populist radical right.
\newblock \emph{West European Politics}, 40\penalty0 (4):\penalty0 791--811,
  2017.


\bibitem[Dandekar et~al.(2013)Dandekar, Goel, and Lee]{Dandekar2013biased}
P.~Dandekar, A.~Goel, and D.~T. Lee.
\newblock Biased assimilation, homophily, and the dynamics of polarization.
\newblock \emph{Proceedings of the National Academy of Sciences}, 110\penalty0
  (15):\penalty0 5791--5796, 2013.

\bibitem[Deffuant et~al.(2000)Deffuant, Neau, Amblard, and
  Weisbuch]{Deffuant2000mixing}
G.~Deffuant, D.~Neau, F.~Amblard, and G.~Weisbuch.
\newblock Mixing beliefs among interacting agents.
\newblock \emph{Advances in Complex Systems}, 3\penalty0 (01n04):\penalty0
  87--98, 2000.

\bibitem[DiMaggio et~al.(1996)DiMaggio, Evans, and Bryson]{DiMaggio1996have}
P.~DiMaggio, J.~Evans, and B.~Bryson.
\newblock Have american's social attitudes become more polarized?
\newblock \emph{American Journal of Sociology}, 102\penalty0 (3):\penalty0
  690--755, 1996.

\bibitem[Dimock et~al.(2014)Dimock, Doherty, Kiley, and
  Oates]{Dimock2014political}
M.~Dimock, C.~Doherty, J.~Kiley, and R.~Oates.
\newblock Political polarization in the american public.
\newblock \emph{Pew Research Center}, 2014.

\bibitem[Downs(1957)]{Downs1957economic}
A.~Downs.
\newblock An economic theory of political action in a democracy.
\newblock \emph{Journal of Political Economy}, 65\penalty0 (2):\penalty0
  135--150, 1957.

\bibitem[Duggins(2017)]{Duggins2017psycologically}
P.~Duggins.
\newblock A psychologically-motivated model of opinion change with applications
  to american politics.
\newblock \emph{Journal of Artificial Societies and Social Simulation},
  20\penalty0 (1):\penalty0 13, 2017.


\bibitem[Eagly and Chaiken(1998)]{Eagly1998attitude}
A.~H. Eagly and S.~Chaiken.
\newblock Attitude structure and function.
\newblock 1998.

\bibitem[Farrar et~al.(2010)Farrar, Fishkin, Green, List, Luskin, and
  Paluck]{Farrar2010disaggregating}
C.~Farrar, J.~S. Fishkin, D.~P. Green, C.~List, R.~C. Luskin, and E.~L. Paluck.
\newblock Disaggregating deliberation’s effects: An experiment within a
  deliberative poll.
\newblock \emph{British Journal of Political Science}, 40\penalty0
  (2):\penalty0 333--347, 2010.

\bibitem[Fishbein(1963)]{Fishbein1963investigation}
M.~Fishbein.
\newblock An investigation of the relationship between beliefs about an object
  and the attitude toward that object.
\newblock \emph{Human Relations}, 1963.

\bibitem[Fishbein and Raven(1962)]{Fishbein1962ab}
M.~Fishbein and B.~H. Raven.
\newblock The AB scales: An operational definition of belief and attitude.
\newblock \emph{Human Relations}, 1962.

\bibitem[Flache and Macy(2011)]{Flache2011small}
A.~Flache and M.~W. Macy.
\newblock Small worlds and cultural polarization.
\newblock \emph{The Journal of Mathematical Sociology}, 35\penalty0
  (1-3):\penalty0 146--176, 2011.

\bibitem[Flache et~al.(2017)Flache, M\"{a}s, Feliciani, Chattoe-Brown,
  Deffuant, Huet, and Lorenz]{Flache2017models}
A.~Flache, M.~M\"{a}s, T.~Feliciani, E.~Chattoe-Brown, G.~Deffuant, S.~Huet,
  and J.~Lorenz.
\newblock Models of social influence: Towards the next frontiers.
\newblock \emph{Journal of Artificial Societies and Social Simulation},
  20\penalty0 (4):\penalty0 2, 2017.

\bibitem[Fortunato et~al.(2005)Fortunato, Latora, Pluchino, and
  Rapisarda]{Fortunato2005vector}
S.~Fortunato, V.~Latora, A.~Pluchino, and A.~Rapisarda.
\newblock Vector opinion dynamics in a bounded confidence consensus model.
\newblock \emph{International Journal of Modern Physics C}, 16\penalty0
  (10):\penalty0 1535--1551, 2005.

\bibitem[Friedkin(2015)]{Friedkin2015problem}
N.~E. Friedkin.
\newblock The problem of social control and coordination of complex systems in
  sociology: A look at the community cleavage problem.
\newblock \emph{IEEE Control Systems}, 35\penalty0 (3):\penalty0 40--51, 2015.

\bibitem[Friedkin et~al.(2016)Friedkin, Proskurnikov, Tempo, and
  Parsegov]{Friedkin2016network}
N.~E. Friedkin, A.~V. Proskurnikov, R.~Tempo, and S.~E. Parsegov.
\newblock Network science on belief system dynamics under logic constraints.
\newblock \emph{Science}, 354\penalty0 (6310):\penalty0 321--326, 2016.

\bibitem[Galam(2005)]{Galam2005local}
S.~Galam.
\newblock Local dynamics vs. social mechanisms: A unifying frame.
\newblock \emph{Europhysics Letters}, 70\penalty0 (6):\penalty0 705--711, 2005.

\bibitem[Goldberg(2011)]{Goldberg2011mapping}
A.~Goldberg.
\newblock Mapping shared understandings using relational class analysis: The
  case of the cultural omnivore reexamined.
\newblock \emph{American Journal of Sociology}, 116\penalty0 (5):\penalty0
  1397--1436, 2011.

\bibitem[Hegselmann et~al.(2002)Hegselmann, Krause,
  et~al.]{Hegselmann2002opinion}
R.~Hegselmann, U.~Krause, et~al.
\newblock Opinion dynamics and bounded confidence models, analysis, and
  simulation.
\newblock \emph{Journal of Artificial Societies and Social Simulation},
  5\penalty0 (3), 2002.

\bibitem[Huet et~al.(2008)Huet, Deffuant, and Jager]{Huet2008rejection}
S.~Huet, G.~Deffuant, and W.~Jager.
\newblock A rejection mechanism in 2d bounded confidence provides more
  conformity.
\newblock \emph{Advances in Complex Systems}, 11\penalty0 (04):\penalty0
  529--549, 2008.

\bibitem[Huston and Levinger(1978)]{Huston1978interpersonal}
T.~L. Huston and G.~Levinger.
\newblock Interpersonal attraction and relationships.
\newblock \emph{Annual Review of Psychology}, 29\penalty0 (1):\penalty0
  115--156, 1978.

\bibitem[Isenberg(1986)]{Isenberg1986group}
D.~J. Isenberg.
\newblock Group polarization: A critical review and meta-analysis.
\newblock \emph{Journal of Personality and Social Psychology}, 50\penalty0
  (6):\penalty0 1141, 1986.

\bibitem[Kobayashi(2016)]{Kobayashi2016relational}
K.~Kobayashi.
\newblock Relational processing of conflicting arguments: Effects on biased
  assimilation.
\newblock \emph{Comprehensive Psychology}, 5:\penalty0 2165222816657801, 2016.

\bibitem[Kondrashov(1986)]{Kondrashov1986multilocus}
A.~S. Kondrashov.
\newblock Multilocus model of sympatric speciation. iii. computer simulations.
\newblock \emph{Theoretical Population Biology}, 29\penalty0 (1):\penalty0
  1--15, 1986.

\bibitem[Kondrashov and Shpak(1998)]{Kondrashov1998origin}
A.~S. Kondrashov and M.~Shpak.
\newblock On the origin of species by means of assortative mating.
\newblock \emph{Proc. R. Soc. Lond. B}, 265:\penalty0 2273--2278, 1998.

\bibitem[Kron(2004)]{Kron2004general}
T.~Kron.
\newblock General theory of action? Inkonsistenzen in der Handlungstheorie von
  Hartmut Esser.
\newblock \emph{Zeitschrift f{\"u}r Soziologie}, 33\penalty0 (3):\penalty0
  186--205, 2004.

\bibitem[Kroneberg(2005)]{Kroneberg2005definition}
C.~Kroneberg.
\newblock Die Definition der Situation und die variable Rationalit{\"a}t der
  Akteure. ein allgemeines Modell des Handelns.
\newblock \emph{Zeitschrift f{\"u}r Soziologie}, 34\penalty0 (5):\penalty0
  344--363, 2005.

\bibitem[Kunda(1990)]{Kunda1990case}
Z.~Kunda.
\newblock The case for motivated reasoning.
\newblock \emph{Psychological Bulletin}, 108\penalty0 (3):\penalty0 480, 1990.

\bibitem[Laver(2014)]{Laver2014measuring}
M.~Laver.
\newblock Measuring policy positions in political space.
\newblock \emph{Annual Review of Political Science}, 17\penalty0 (1):\penalty0
  207--223, 2014.

\bibitem[Lazarsfeld and Merton(1954)]{Lazarsfeld1954friendship}
P.~Lazarsfeld and R.~K. Merton.
\newblock Friendship as a social process: A substantive and methodological
  analysis.
\newblock In M.~Berger, T.~Abel, and C.~H. Page, editors, \emph{Freedom and
  Control in Modern Society}, pages 18--66. New York: Van Nostrand, 1954.

\bibitem[Leuthold et~al.(2007)Leuthold, Hermann, and
  Fabrikant]{Leuthold2007making}
H.~Leuthold, M.~Hermann, and S.~I. Fabrikant.
\newblock Making the political landscape visible: Mapping and analyzing voting
  patterns in an ideological space.
\newblock \emph{Environment and Planning B: Planning and Design}, 34\penalty0
  (5):\penalty0 785--807, 2007.

\bibitem[Lindenberg(1992)]{Lindenberg1992method}
S.~Lindenberg.
\newblock The method of decreasing abstraction.
\newblock \emph{Rational Choice Theory: Advocacy and Critique}, 1:\penalty0 6,
  1992.

\bibitem[Lippi and Torroni(2016)]{Lippi2016argumentation}
M.~Lippi and P.~Torroni.
\newblock Argumentation mining: State of the art and emerging trends.
\newblock \emph{ACM Trans. Internet Technol.}, 16\penalty0 (2):\penalty0
  10:1--10:25, Mar. 2016.


\bibitem[Lord et~al.(1979)Lord, Ross, and Lepper]{Lord1979biased}
C.~G. Lord, L.~Ross, and M.~R. Lepper.
\newblock Biased assimilation and attitude polarization: The effects of prior
  theories on subsequently considered evidence.
\newblock \emph{Journal of Personality and Social Psychology}, 37\penalty0
  (11):\penalty0 2098, 1979.

\bibitem[Lorenz(2003)]{Lorenz2003mehrdimensionale}
J.~Lorenz.
\newblock Mehrdimensionale meinungsdynamik bei wechselndem vertrauen.
\newblock \emph{Master's thesis, University of Bremen}, 2003.

\bibitem[Lorenz(2007)]{Lorenz2007continuous2}
J.~Lorenz.
\newblock Continuous opinion dynamics under bounded confidence: A survey.
\newblock \emph{International Journal of Modern Physics C}, 18\penalty0
  (12):\penalty0 1819--1838, 2007.

\bibitem[Lorenz(2008)]{Lorenz2008fostering}
J.~Lorenz.
\newblock Fostering consensus in multidimensional continuous opinion dynamics
  under bounded confidence.
\newblock In \emph{Managing complexity: insights, concepts, applications},
  pages 321--334. Springer, 2008.

\bibitem[Macy et~al.(2003)Macy, Kitts, Flache, and
  Benard]{Macy2003polarization}
M.~W. Macy, J.~A. Kitts, A.~Flache, and S.~Benard.
\newblock Polarization in dynamic networks: A hopfield model of emergent
  structure.
\newblock \emph{Dynamic social network modeling and analysis}, pages 162--173,
  2003.

\bibitem[M{\"a}s and Bischofberger(2015)]{Maes2015will}
M.~M{\"a}s and L.~Bischofberger.
\newblock Will the personalization of online social networks foster opinion
  polarization?
\newblock \emph{Available at SSRN 2553436}, 2015.

\bibitem[M{\"a}s and Flache(2013)]{Maes2013differentiation}
M.~M{\"a}s and A.~Flache.
\newblock Differentiation without distancing. explaining bi-polarization of
  opinions without negative influence.
\newblock \emph{PloS one}, 8\penalty0 (11):\penalty0 e74516, 2013.

\bibitem[M{\"a}s et~al.(2010)M{\"a}s, Flache, and
  Helbing]{Maes2010individualization}
M.~M{\"a}s, A.~Flache, and D.~Helbing.
\newblock Individualization as driving force of clustering phenomena in humans.
\newblock \emph{PLoS Computational Biology}, 6\penalty0 (10):\penalty0
  e1000959, 2010.

\bibitem[M{\"a}s et~al.(2013)M{\"a}s, Flache, Tak{\'a}cs, and
  Jehn]{Maes2013short}
M.~M{\"a}s, A.~Flache, K.~Tak{\'a}cs, and K.~A. Jehn.
\newblock In the short term we divide, in the long term we unite: Demographic
  crisscrossing and the effects of faultlines on subgroup polarization.
\newblock \emph{Organization Science}, 24\penalty0 (3):\penalty0 716--736,
  2013.

\bibitem[Mason(2018)]{Mason2018uncivil}
L.~Mason.
\newblock \emph{Uncivil agreement: How politics became our identity}.
\newblock University of Chicago Press, 2018.

\bibitem[McPherson et~al.(2001)McPherson, Smith-Lovin, and
  Cook]{McPherson2001birds}
M.~McPherson, L.~Smith-Lovin, and J.~M. Cook.
\newblock Birds of a feather: Homophily in social networks.
\newblock \emph{Annual Review of Sociology}, 27:\penalty0 415--444, 2001.

\bibitem[Mueller and Tan(2018)]{Mueller2018cognitive}
S.~T. Mueller and Y.-Y.~S. Tan.
\newblock Cognitive perspectives on opinion dynamics: the role of knowledge in
  consensus formation, opinion divergence, and group polarization.
\newblock \emph{Journal of Computational Social Science}, 1\penalty0
  (1):\penalty0 15--48, 2018.

\bibitem[Parsegov et~al.(2016)Parsegov, Proskurnikov, Tempo, and
  Friedkin]{Parsegov2016novel}
S.~E. Parsegov, A.~V. Proskurnikov, R.~Tempo, and N.~E. Friedkin.
\newblock Novel multidimensional models of opinion dynamics in social networks.
\newblock \emph{IEEE Transactions on Automatic Control}, 2016.

\bibitem[Rosa(2016)]{Rosa2016Resonanz}
H.~Rosa.
\newblock \emph{Resonanz: Eine Soziologie der Weltbeziehung}.
\newblock Suhrkamp Verlag, 2016.

\bibitem[Schütz and Luckmann(2017)]{Schuetz2017strukturen}
A.~Schütz and T.~Luckmann.
\newblock \emph{Strukturen der Lebenswelt}, volume 2. überarbeitete Auflage.
\newblock UVK, Konstanz u. München, 2017.

\bibitem[Shamon et~al.(forthcoming)Shamon, Schumann, Fischer, Vögele,
  Heinrichs, and Kuckshinrichs]{Shamon2019arguments}
H.~Shamon, D.~Schumann, W.~Fischer, S.~Vögele, H.~U. Heinrichs, and
  W.~Kuckshinrichs.
\newblock Arguments on electricity generating technologies: Examining biased
  processing, attitude polarization and awareness effects.
\newblock forthcoming.

\bibitem[Sobkowicz(2009)]{Sobkowicz2009modelling}
P.~Sobkowicz.
\newblock Modelling opinion formation with physics tools: Call for closer link
  with reality.
\newblock \emph{Journal of Artificial Societies and Social Simulation},
  12\penalty0 (1):\penalty0 11, 2009.

\bibitem[Steels(2017)]{Steels2017basics}
L.~Steels.
\newblock Basics of fluid construction grammar.
\newblock \emph{Constructions and Frames}, 9\penalty0 (2):\penalty0 178--225,
  2017.

\bibitem[Strauss and Quinn(1997)]{Strauss1997cognitive}
C.~Strauss and N.~Quinn.
\newblock \emph{A cognitive theory of cultural meaning}, volume~9.
\newblock Cambridge University Press, 1997.

\bibitem[Sunstein(2002)]{Sunstein2002law}
C.~R. Sunstein.
\newblock The law of group polarization.
\newblock \emph{Journal of Political Philosophy}, 10\penalty0 (2):\penalty0
  175--195, 2002.

\bibitem[Sznajd-Weron and Sznajd(2000)]{Sznajd-Weron2000opinion}
K.~Sznajd-Weron and J.~Sznajd.
\newblock Opinion evolution in closed community.
\newblock \emph{International Journal of Modern Physics C}, 11:\penalty0
  1157--1165, 2000.

\bibitem[Taber et~al.(2009)Taber, Cann, and Kucsova]{Taber2009motivated}
C.~S. Taber, D.~Cann, and S.~Kucsova.
\newblock The motivated processing of political arguments.
\newblock \emph{Political Behavior}, 31\penalty0 (2):\penalty0 137--155, 2009.

\bibitem[Uitermark et~al.(2016)Uitermark, Traag, and
  Bruggeman]{Uitermark2016dissecting}
J.~Uitermark, V.~A. Traag, and J.~Bruggeman.
\newblock Dissecting discursive contention: A relational analysis of the dutch
  debate on minority integration, 1990--2006.
\newblock \emph{Social Networks}, 47:\penalty0 107--115, 2016.

\bibitem[Urbig and Malitz(2005)]{Urbig2005dynamics}
D.~Urbig and R.~Malitz.
\newblock Dynamics of structured attitudes and opinions.
\newblock In \emph{Third Conference of the European Social Simulation
  Association}, pages 5--8. Citeseer, 2005.

\bibitem[{Van Eecke} and Beuls(2018)]{VanEecke2018exploring}
P.~{Van Eecke} and K.~Beuls.
\newblock Exploring the creative potential of computational construction
  grammar.
\newblock \emph{Zeitschrift f{\"u}r Anglistik und Amerikanistik}, 66\penalty0
  (3):\penalty0 341--355, 2018.

\bibitem[Van~Overwalle and Heylighen(2006)]{VanOverwalle2006talking}
F.~Van~Overwalle and F.~Heylighen.
\newblock Talking nets: A multiagent connectionist approach to communication
  and trust between individuals.
\newblock \emph{Psychological Review}, 113\penalty0 (3):\penalty0 606, 2006.

\bibitem[Van~Overwalle and Siebler(2005)]{VanOverwalle2005connectionist}
F.~Van~Overwalle and F.~Siebler.
\newblock A connectionist model of attitude formation and change.
\newblock \emph{Personality and Social Psychology Review}, 9\penalty0
  (3):\penalty0 231--274, 2005.

\bibitem[Wimmer and Lewis(2010)]{Wimmer2010beyond}
A.~Wimmer and K.~Lewis.
\newblock Beyond and below racial homophily: Erg models of a friendship network
  documented on facebook.
\newblock \emph{American Journal of Sociology}, 116\penalty0 (2):\penalty0
  583--642, 2010.

\bibitem[Wolf et~al.(2012)Wolf, Neumann, Schr{\"o}der, and
  de~Haan]{Wolf2012adoption}
I.~Wolf, J.~Neumann, T.~Schr{\"o}der, and G.~de~Haan.
\newblock The adoption of electric vehicles: An empirical agent-based model of
  attitude formation and change.
\newblock In \emph{Proceedings of the 8th Conference of the European
  Association for Social Simulation}, pages 93--98, 2012.

\bibitem[Wolf et~al.(2015)Wolf, Schr{\"o}der, Neumann, and
  de~Haan]{Wolf2015changing}
I.~Wolf, T.~Schr{\"o}der, J.~Neumann, and G.~de~Haan.
\newblock Changing minds about electric cars: An empirically grounded
  agent-based modeling approach.
\newblock \emph{Technological Forecasting and Social Change}, 94:\penalty0
  269--285, 2015.

\end{thebibliography}

\end{document}